\documentclass[
reprint,
 amsmath,amssymb,
 aps,twocolumn,
]{revtex4-2}

\usepackage{longtable}
\usepackage{graphicx}
\usepackage{dcolumn}
\usepackage{bm}
\usepackage[colorlinks,
linkcolor=black,anchorcolor=black,
citecolor=black,urlcolor=black]{hyperref}
\usepackage{multirow}
\usepackage{diagbox} 
\usepackage{color}
\usepackage{longtable}
\usepackage{supertabular}
\usepackage{amsmath} 
\usepackage{array, booktabs, ulem}
\allowdisplaybreaks[4] 
\usepackage{bm}
\usepackage{graphicx}
\usepackage{subfigure} 

\usepackage{xurl}

\begin{document}

\preprint{APS/123-QED}

\title{Effects of lunisolar perturbations on TianQin constellation: An analytical model}


\author{Bobing Ye}
 \email{yebb5@mail.sysu.edu.cn}
\author{Xuefeng Zhang}
 \email{zhangxf38@sysu.edu.cn}
\affiliation{ 
MOE Key Laboratory of TianQin Mission, TianQin Research Center for
Gravitational Physics\\ $ \& $ School of Physics and Astronomy, Frontiers Science Center for TianQin, \\Gravitational Wave Research Center of CNSA, Sun Yat-sen University (Zhuhai Campus), Zhuhai 519082, People's Republic of China 
}


\date{\today}

\begin{abstract}

TianQin is a proposed space-based gravitational-wave observatory mission that critically relies on the stability of an equilateral-triangle constellation. Comprising three satellites in high Earth orbits of a $ 10^5 $ km radius, this constellation's geometric configuration is significantly affected by gravitational perturbations, primarily originating from the Moon and the Sun. In this paper, we present an analytical model to quantify the effects of lunisolar perturbations on the TianQin constellation, derived using Lagrange's planetary equations. The model provides expressions for three kinematic indicators of the constellation: arm-lengths, relative line-of-sight velocities, and breathing angles. Analysis of these indicators reveals that lunisolar perturbations can distort the constellation triangle, resulting in three distinct variations: linear drift, bias, and fluctuation. Furthermore, it is shown that these distortions can be optimized to display solely fluctuating behavior, under certain predefined conditions. These results can serve as the theoretical foundation for numerical simulations and offer insights for engineering a stable constellation in the future.

\end{abstract}

\maketitle



\section{\label{sec:level1} Introduction}

The successful detection of gravitational waves (GWs) by the ground-based observatory LIGO \cite{Abbott2016} has opened up the era of GW astronomy. To detect GWs in the millihertz range (0.1 mHz–1 Hz), known for its rich sources and to circumvent the impact of seismic noise, space-based GW observatories are highly favored \cite{Ni2016,NASA2012}. For such observatories, proposed projects include LISA \cite{LISA2017,LISA2024}, DECIGO \cite{Seto2001}, TianQin \cite{Luo2016}, Taiji \cite{HuWR2017}, etc. Among these, TianQin is a geocentric space-based GW observatory mission that consists of three drag-free controlled satellites with an orbital radius of $ 10^{5} $ km \cite{Luo2016}. The three satellites form a nearly equilateral-triangle constellation, standing almost vertical to the ecliptic, and they employ high-precision laser-ranging interferometry to measure distance changes between satellites for the detection of GWs. The mission will bring rich science prospects to GW astronomy \cite{Hu2017,Mei2020,Gong2021}.

TianQin, as well as other three-satellite GW missions, relies crucially on the stability of an equilateral-triangle constellation \cite{NASA2012,Luo2016}. Unequal variations in the three arm-lengths of the constellation prevent the cancellation of laser frequency noise, which has a profound impact on the design of frequency stabilization systems and requires time-delay interferometry (TDI)  \cite{Tinto2014,Zhou2021,Zheng2023}. The relative line-of-sight velocities between satellites induce Doppler frequency shifts, affecting phase meter bandwidth and ultra-stable oscillator design \cite{Folkner1997}. Moreover, changes in the three breathing angles of
the triangle directly influence the design of telescopes and beam pointing mechanisms \cite{Luo2016}. It is crucial to minimize variations in the triangular constellation, as indicated by these three kinematic indicators.

Analytical analysis of satellite motion and constellation variations holds significant importance \cite{Roscoe2013,Wu2019,deMarchi2012,Qiao2023b,Dhurandhar2005,Wu2019}. To identify orbits with minimal variations in the constellation, extensive efforts have been dedicated to numerical orbit optimization and analysis (for a review, see Ref.~\cite{Qiao2023b}). The use of analytical models, as opposed to numerical simulations, allows for deeper physical insights and often yields clearer solutions for issues related to satellite motion \cite{Roscoe2013,Wu2019}. Moreover, these analytical models provide the basis for further numerical simulations, enhancing orbit optimization efficiency \cite{deMarchi2012,Qiao2023b}. They also enable theoretical studies on inter-satellite optical links and light propagation \cite{Dhurandhar2005,Wu2019}.

Concerning analytical efforts, Ref.~\cite{Hu2018} first presented the analytical coordinates of the TianQin satellites, based on unperturbed Keplerian orbits, which showed that the arm-lengths of the constellation remain constant when orbital eccentricities are ignored. Furthermore, the leading-order effect of the third-body perturbation was considered to derive expressions for both arm-lengths and breathing angles \cite{Qiao2023a}. These expressions were constructed iteratively, assuming circular orbits, and they were used to study the impact of initial orbit errors. Moreover, the effect of the Earth's non-spherical gravitational perturbation was analyzed in \cite{Jiao2023}, with a particular focus on its influence on inter-satellite range acceleration noise.

The analytical investigation into the influence of gravitational perturbations on the TianQin constellation is incomplete. Existing models have neglected the satellite's orbital eccentricity, a crucial factor for constellation stability \cite{Ye2019, Tan2020}. Moreover, relying solely on the leading-order lunar perturbation is insufficient to address the high-altitude TianQin orbits. These issues highlight the necessity for an analytical study to develop a more explicit and higher-precision model.

In the exploration of three-satellite constellations in heliocentric GW missions, such as LISA \cite{Martens2021, Dhurandhar2005, Nayak2006, Yi2008, Cerdonio2010CQG, Pucacco2010, deMarchi2012} and Taiji \cite{Wu2019, Wu2020}, expressions for these three indicators have been derived and analyzed using either Keplerian orbits or perturbation solutions of satellite orbits. Valuable references are also found in geocentric satellite formation missions, including NASA's four-satellite Magnetospheric Multiscale (MMS) mission \cite{Roscoe2013}, and extensive studies on third-body perturbations in general satellites (see \cite{Nie2019} and references therein). Perturbation solutions for third-body effects can be derived by solving Lagrange's planetary equations \cite{Capderou2014}, where the perturbative potential depends on the orbital elements of both the satellite and perturbing bodies. To directly obtain solutions with instantaneous elements, perturbation methods \cite{Capderou2014, Kozai1973, Liu2000, Xu2011}, especially the mean element method \cite{Kozai1973, Liu2000}, are utilized. This method employs a slowly precessing elliptical orbit as a reference, effectively reducing errors in analytical solutions.

In this work, we will construct an analytical model for the TianQin constellation. To address its near-circular, high Earth orbits, we utilize singularity-free Lagrange equations while accounting for lunar, solar perturbations, and Earth's $ J_{2} $ perturbation. This model will then be used to analyze and optimize the three kinematic indicators. Additionally, to facilitate the perturbation-inclusive study, the unperturbed Keplerian orbits of TianQin satellites will also be presented.

The paper is organized as follows. In Sec.~\ref{sec:2}, we introduce the Keplerian orbits of the satellites and present the design of the nominal equilateral-triangle constellation. The gravitational perturbations on the constellation are studied in Sec.~\ref{sec:3}. In Sec.~\ref{sec:4}, we make the concluding remarks.


\section{\label{sec:2} Fundamentals of stable TianQin constellation}

In this section, we describe the motion of TianQin satellites in the geocentric ecliptic coordinate system and present the orbit design of satellites for a stable equilateral-triangle constellation.

\subsection{\label{sec:2.1} Keplerian orbits of satellites}

\begin{figure}[htb]
\includegraphics[width=0.46\textwidth]{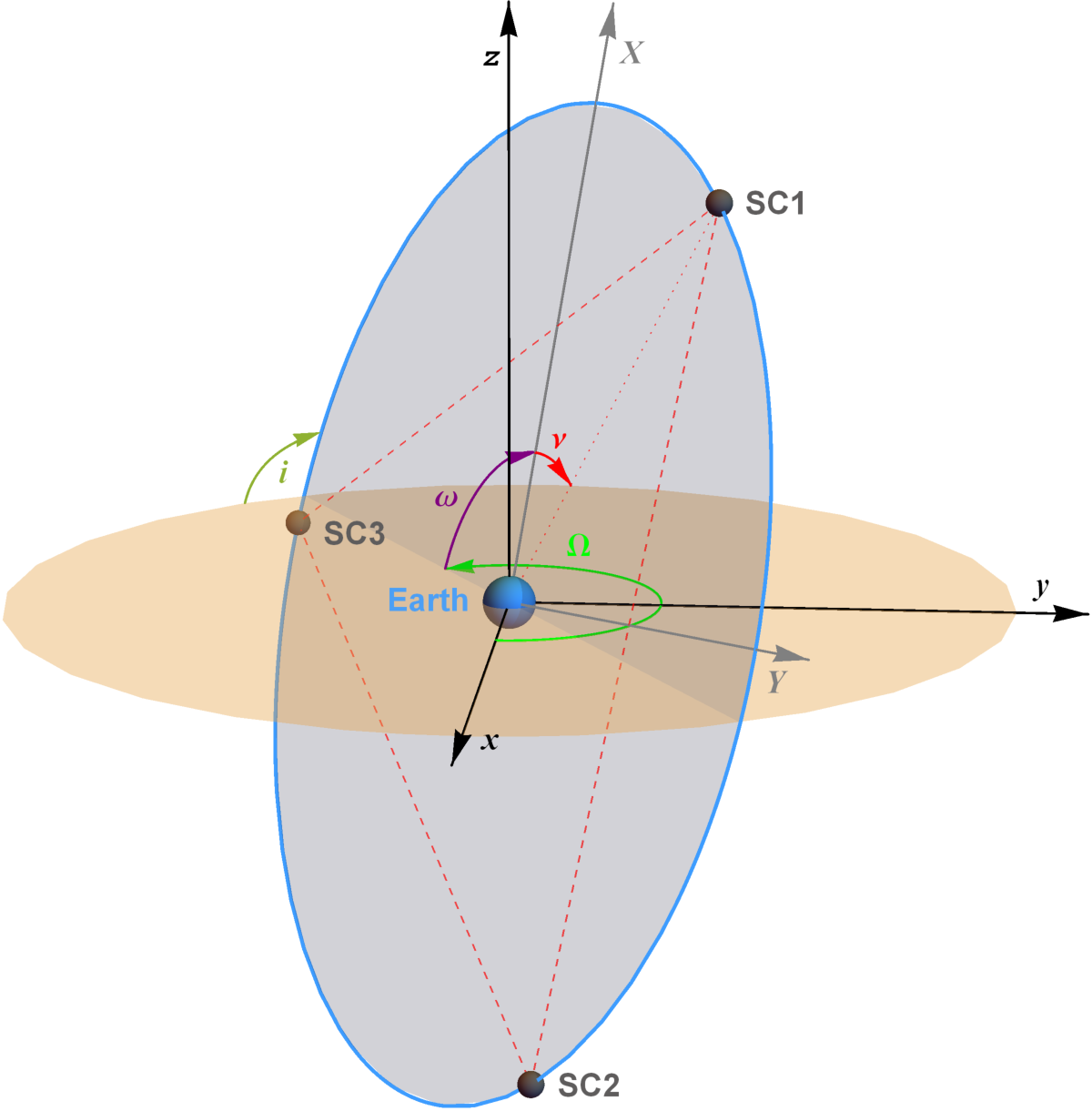}
\caption{\label{fig:orbit} Depiction of the TianQin constellation in the geocentric ecliptic coordinate system. The ecliptic plane is spanned by the $ x $ and $ y $ axes, with the $ x $-axis directed toward the vernal equinox. The figure also illustrates the orbital coordinate system $\{X, Y, Z\}$ for SC1, where the $ X $-axis points toward the perigee of the satellite's orbit, and the $ Z $-axis (not shown) is perpendicular to the orbital plane. The angles $ i $, $ \Omega $, $ \omega $, and $ \nu $ denote the orbital inclination, longitude of ascending node, argument of perigee, and true anomaly, respectively. Specifically, $ i = 94.7^\circ $ and $ \Omega = 210.4^\circ $ are set to orient the TianQin detector plane toward the reference source, the white-dwarf binary RX J0806.3$\texttt{+} $1527.}
\end{figure}

Within the central gravitational field of the Earth, satellite moves in a Keplerian orbit, as illustrated in Fig.~\ref{fig:orbit}.
$\{X, Y, Z\}$ is the orbital right-handed coordinate system, with the origin at the Earth's center of mass. The satellite's orbital plane is same as the $X$-$Y$ plane, where the $X$-axis points toward the perigee. In this system, the satellite's Cartesian coordinates $(X, Y, Z, \dot{X}, \dot{Y}, \dot{Z})$ can be denoted as \cite{Liu2000,Capderou2014}: 
\begin{align}
\left\{
\begin{array}{l}
X = r \cos \nu =  a(\cos E-e),\smallskip\\
Y = r \sin \nu = a\sqrt{1-e^{2}} \sin E,\smallskip\\
Z = 0,\smallskip\\
\dot{X} = -\frac{\sqrt{\mu a}}{r} \sin E,\smallskip\\
\dot{Y} = \frac{\sqrt{\mu a}}{r} \sqrt{1-e^{2}} \cos E, \smallskip\\
\dot{Z} = 0,
\end{array} 
\right.\label{eq:XYZ}
\end{align}
with $r$ representing the geocentric radius, $\nu$  the true anomaly, $a$ the semimajor axis, $e$ the orbital eccentricity, $\mu=G M_{\text{e}}$ the Earth's gravitational constant, and $E$ the eccentric anomaly. $ E $ satisfies Kepler's equation,
\begin{align}
E - e\sin E= M,\label{eq:EM}
\end{align}
where $ M $ denotes the mean anomaly. Specifically, $ M $ is given by $ M = n\, (t - t_{\text{p}}) $ in the two-body problem, with  the mean motion $ n $ and the passing time of the perigee $ t_{\text{p}} $. Equation (\ref{eq:EM}), which is a transcendental equation, can be solved iteratively, resulting in the following expression \cite{Hu2018}:
\begin{align}
E= M +e \sin M+e^{2}\cos M \sin M+\mathcal{O}(e^{3}). \label{eq:E}
\end{align}
By substituting Eq.~(\ref{eq:E}) into Eq.~(\ref{eq:XYZ}), one can obtain the explicit coordinates $(X, Y, Z, \dot{X}, \dot{Y}, \dot{Z})$.

The orbital planes may not be identical for the three TianQin satellites. Thus, the geocentric ecliptic coordinate system $ \{x, y, z\} $ is also employed, where the $ x$-$y $ plane is the ecliptic plane. The $x$-axis points toward the vernal equinox, and the $z$-axis is normal to the ecliptic plane. The coordinates $ (x, y, z) $ and $ (\dot{x}, \dot{y}, \dot{z}) $ in this system can be obtained  by $(X, Y, Z) $ and $ (\dot{X}, \dot{Y}, \dot{Z}) $ through the following transformation \cite{Liu2000,Capderou2014}: 
\begin{align}
\begin{bmatrix}
x \\ 
y \\ 
z
\end{bmatrix} 
= R_{z}(-\Omega)R_{x}(-i)R_{z}(- \omega ) 
\begin{bmatrix}
X \\ 
Y \\ 
Z
\end{bmatrix},\label{eq:xyz}
\\
\begin{bmatrix}
\dot{x} \\ 
\dot{y} \\ 
\dot{z}
\end{bmatrix} 
= R_{z}(-\Omega)R_{x}(-i)R_{z}(- \omega ) 
\begin{bmatrix}
\dot{X} \\ 
\dot{Y} \\ 
\dot{Z}
\end{bmatrix},\label{eq:Vxyz}
\end{align}
where $\Omega$, $i$, and $\omega$ denote the satellite's longitude of the ascending node, inclination, and argument of perigee, respectively. Additionally, $R_z(\gamma)$ and $R_x(\gamma)$ are the rotation matrices that rotate vectors by an angle $  \gamma $ about the $ z-$ or $ x-$axis,
\begin{align}
R_{z}(\gamma ) = \begin{bmatrix}
\cos \gamma  & \sin \gamma   & 0\\ 
- \sin \gamma   & \cos \gamma  & 0\\ 
0 &  0 & 1
\end{bmatrix}, \label{eq:matrix1}
 \\
R_{x}(\gamma ) = \begin{bmatrix}
1 & 0 & 0\\ 
0 & \cos \gamma  & \sin \gamma  \\ 
0 & - \sin \gamma  & \cos \gamma 
\end{bmatrix}.\label{eq:matrix2}
\end{align}
Combining Eqs.~(\ref{eq:XYZ}) and (\ref{eq:E})-(\ref{eq:matrix2}), the position vector and velocity vector of SC$ k $ ($ k = 1 $, 2, 3), $ \mathbf{r}_{k} = (x_{k}, y_{k}, z_{k}) $ and $ \mathbf{\dot{r}}_{k} = (\dot{x}_{k}, \dot{y}_{k}, \dot{z}_{k}) $, are given by
\begin{align}\label{eq:z_k}
\left\{\begin{array}{l}
\displaystyle x_{k} = a_{k} [\cos  \Omega_{k}  \cos  \lambda_{k} -\cos  i_{k} \sin  \Omega_{k}  \sin  \lambda_{k} \smallskip\\
\displaystyle\phantom{x_{k}=} +\frac{1}{2} e_{k} (\cos  \Omega_{k}  \,f_{\text{c1}k} -\cos  i_{k} \sin  \Omega_{k}  \,f_{\text{s1}k})]+\mathcal{O}(e^{2}_{k}), \smallskip\\
\displaystyle y_{k} =a_{k} [\sin  \Omega_{k}  \cos  \lambda_{k} +\cos  i_{k} \cos  \Omega_{k}  \sin  \lambda_{k} \smallskip\\
\displaystyle\phantom{y_{k}=} +\frac{1}{2} e_{k} (\sin  \Omega_{k}  \,f_{\text{c1}k}+\cos  i_{k} \cos  \Omega_{k}   \,f_{\text{s1}k})]+\mathcal{O}(e^{2}_{k}),   \smallskip\\
\displaystyle z_{k} =a_{k} (\sin  i_{k} \sin  \lambda_{k} +\frac{1}{2} e_{k} \sin  i_{k} \, f_{\text{s1}k} )+\mathcal{O}(e^{2}_{k}),  \smallskip\\
\displaystyle \dot{x}_{k} = \frac{\sqrt{\mu}}{\sqrt{a_{k}}} [-\cos  \Omega _k \sin \lambda _k -\cos  i_k  \sin  \Omega _k \cos  \lambda _k  \smallskip\\
\displaystyle\phantom{\dot{x}_{k}=} +e_k (-\cos  \Omega _k  \, f_{\text{s2}k} -\cos  i_k  \sin  \Omega _k \, f_{\text{c2}k} ) ] +\mathcal{O}(e^{2}_{k}),\smallskip\\
\displaystyle \dot{y}_{k} = \frac{\sqrt{\mu}}{\sqrt{a_{k}}} [- \sin  \Omega _k \sin  \lambda _k  + \cos  i_k \cos  \Omega _k \cos  \lambda _k  \smallskip\\
\displaystyle\phantom{\dot{y}_{k}=}+e_k (- \sin  \Omega _k \, f_{\text{s2}k} + \cos  i_k  \cos  \Omega _k \, f_{\text{c2}k} ) ]+\mathcal{O}(e^{2}_{k}),\smallskip\\
\displaystyle \dot{z}_{k} = \frac{\sqrt{\mu}}{\sqrt{a_{k}}} (\sin  i_k \cos  \lambda _k + e_k \sin  i_k \, f_{\text{c2}k})+\mathcal{O}(e^{2}_{k}), 
\end{array}\right.
\end{align}
where
\begin{align}
\lambda_{k} := M_{k} + \omega_{k},
\end{align}
$ f_{\text{c1}k} := \cos (2 \lambda_{k} -\omega_{k} )-3 \cos  \omega_{k} $, $ f_{\text{s1}k} := \sin (2   \lambda_{k} -\omega_{k} )-3 \sin  \omega_{k} $, $ f_{\text{c2}k} := \cos (2 \lambda_{k} -\omega_{k} ) $, and $ f_{\text{s2}k} := \sin (2   \lambda_{k} -\omega_{k} ) $. Define $ \sigma \in \{ a, e, i, \Omega, \omega, \lambda \} $, $ \sigma_{k}(t) $ in Eq.~(\ref{eq:z_k}) are straightforwardly determined in the two-body problem by
\begin{align} 
\sigma_{k} (t) = \sigma_{0k} +\delta_{\iota} n_{k} (t-t_{0}),\qquad \text{for Keplerian orbit}, \label{eq:sol_unperturbed_sec2}
\end{align}
where
\begin{equation}
\sigma_{0k} := \sigma_{k} (t_{0}), \qquad \delta_{\iota} = \left\{
\begin{array}{ll}
1, & \sigma = \lambda, \smallskip\\
0, & \sigma \neq  \lambda. 
\end{array}
\right. \label{eq:delta_01_a}
\end{equation}
Note that Eq.~(\ref{eq:z_k}) remains valid even when considering gravitational perturbations, with the only change being the replacement of $ \sigma_{k}(t) $ in Eq.~(\ref{eq:sol_unperturbed_sec2}) with the corresponding perturbation solution.

\subsection{Orbit design of the TianQin constellation}

The TianQin constellation is composed of three satellites in geocentric orbits, forming a triangular configuration which continuously evolves in geometry over time. The closer the configuration change approaches an equilateral triangle, the more it aids in alleviating design constraints on measurement system instruments. Therefore, it is essential to find a constellation design with minimal variations.

The constellation is considered more stable if it is closer to an equilateral triangle. There are three main kinematic indicators to characterize the stability, namely, the arm-length  $L_{ij}$, relative line-of-sight velocity between satellites $v_{ij}$, and breathing angle $\alpha_{k}$, 
\begin{align}
L_{ij} &=  \left | \mathbf{r}_{i} - \mathbf{r}_{j} \right |,\label{eq:L_ij}\\
v_{ij} &= \dot{L}_{ij},\label{eq:v_ij}\\
\alpha_{k} &= \arccos \frac{L_{ki}^{2}+L_{kj}^{2}-L_{ij}^{2}}{2 L_{ki} L_{kj}},\label{eq:alpha_k}  
\end{align}
where $i$, $j$, and $k$ take values 1, 2, or 3 and $i \neq j \neq k$. Substituting Eq.~(\ref{eq:z_k}) into Eqs.~(\ref{eq:L_ij})-(\ref{eq:alpha_k}), one can obtain the explicit expressions, for these three kinematic indicators, with forms $ L_{ij}(\sigma_{i}(t),\sigma_{j}(t)) $, $ v_{ij}(\sigma_{i}(t),\sigma_{j}(t)) $, and $ \alpha_{k}(\sigma_{k}(t),\sigma_{i}(t),\sigma_{j}(t)) $, respectively.

To maintain the constellation as an equilateral triangle, i.e. $L_{12}(t)=L_{13}(t)=L_{23}(t)$, the orbits of the three satellites need to be purposefully designed. One intuitive orbit design involves circular orbits for the satellites in the point-mass gravitational field of the Earth: 
\begin{align}  \label{eq:e_0value}
 e_{1}(t) = e_{2}(t) = e_{3}(t) = 0,
\end{align}
while ensuring that they share the same orbit size, lie in the same orbital plane, and are phased 120 degrees apart from each other:
\begin{align}
\left\{
\begin{array}{l}
a_{1}(t) = a_{2}(t) = a_{3}(t),\smallskip \\
i_{1}(t) = i_{2}(t) = i_{3}(t),\smallskip\\
\Omega_{1}(t) = \Omega_{2}(t) = \Omega_{3}(t),\smallskip\\
\lambda_{k}(t) = \frac{2\pi }{3}(k-1) + \lambda_{1}(t).
\end{array} 
\right.\label{eq:NominalOrbit_con}
\end{align}
The above requirements on the inter-satellite parameters can be achieved in the two-body problem, if the initial orbital elements $ \sigma_{0k} $ in Eq.~(\ref {eq:sol_unperturbed_sec2}) are set to 
\begin{align}
\sigma_{0k} = \sigma_{\text{o}} +\delta_{\iota} \frac{2\pi }{3}(k-1),\label{eq:NominalOrbit_con_t0}
\end{align}
where the parameters with subscript ``$\text{o}$" are the nominal ones of the TianQin constellation. For instance, these values can be chosen as $a_{\text{o}} = 10^5$ km, $e_{\text{o}} = 0$, and $i_{\text{o}} = 94.7^\circ$, $ \Omega_{\text{o}} = 210.4^\circ$, respectively establishing the orbit size and orienting the orbital plane perpendicular to J0806 \cite{Luo2016,Ye2019}. The initial value $ \lambda_{\text{o}} $ associated with the orbit phase is typically selected to be any value within the range of $0^\circ$ to $120^\circ$, or it may be specifically designated to avoid Moon eclipses \cite{Ye2021}. 

To analyze additional nominal orbit design allowing for $ e \neq 0 $ and quantify the impact of eccentricity on the three indicators, the constraint specified by Eq.~(\ref{eq:e_0value}) is relaxed. Subsequently, employing only Eq.~(\ref{eq:NominalOrbit_con}) or Eq.~(\ref{eq:NominalOrbit_con_t0}) (for $ \sigma \in \{a, i, \Omega, \lambda \} $), the variations of these indicators in the two-body problem, up to the first order of $ e $, can be expressed as 
\begin{align}
 L_{ij}^{\text{kepl}}(t)= &~ \sqrt{3} a_{\text{o}}+\frac{\sqrt{7}}{2}  a_{\text{o}} [-e_i \sin \left(M_i+\beta\right) \nonumber\\
& +e_j \sin \left(M_j-\beta\right)], \label{eq:L_ij_e1} 
\end{align}
\begin{align}
v_{ij}^{\text{kepl}}(t)= &~ 0+\frac{\sqrt{7}}{2}  a_{\text{o}} n_{\text{o}} [-e_i \cos \left(M_i+\beta\right) \nonumber\\
& +e_j \cos \left(M_j-\beta\right)],\label{eq:v_ij_e1}
\end{align}
\begin{align}
\alpha _k^{\text{kepl}}(t)= &~ \frac{\pi }{3}+\frac{ \sqrt{7}}{3} e_k \cos M_k \sin \beta  \nonumber\\
& -\frac{\sqrt{7}}{6}  e_i  [\sin \left(M_i-\beta\right)+2 \sin \left(M_i+\beta\right)] \nonumber \\
&  +\frac{\sqrt{7}}{6}  e_j [2 \sin   \left(M_j-\beta\right)+\sin \left(M_j+\beta\right)],\label{eq:alpha_k_e1}  
\end{align} 
where $ M_{i}(t) = \frac{2\pi }{3}(i-1) + \lambda_{1}(t) - \omega_{i}  $ and $ \beta := \arccos \frac{2}{\sqrt{7}} $, with the indices $i$, $j$, and $k$ using cyclic indexing ($ i, \,j,\, k = 1 \to 2 \to 3 \to 1$). If we further set $e_{1} = e_{2} = e_{3} \equiv e_{\text{o}} \neq 0 $ and $ \omega_{1} = \omega_{2} = \omega_{3} $, then it follows that 
\begin{align}
& L_{ij}^{\text{kepl-A}}(t)=\sqrt{3} a_{\text{o}}-\frac{\sqrt{3}}{2}  a_{\text{o}}  e_{\text{o}} \cos M_{k} ,\label{eq:L_ij_ek=e1}  \\
& v_{ij}^{\text{kepl-A}}(t)= 0 + \frac{\sqrt{3}}{2}  a_{\text{o}} n_{\text{o}} e_{\text{o}} \sin M_{k} ,\label{eq:v_ij_ek=e1} \\
& \alpha _k^{\text{kepl-A}}(t)=\frac{\pi }{3}-\frac{\sqrt{3}}{2} e_{\text{o}} \cos M_{k}
,\label{eq:alpha_k_ek=e1}  
\end{align}
where $ M_{k}(t) = \frac{2\pi }{3}(k-1) + M_{1}(t) $.
Equations (\ref{eq:L_ij_ek=e1})-(\ref{eq:alpha_k_ek=e1}) indicate that, at the zeroth order of $ e $, the three TianQin satellites can form a constant equilateral triangle. However, when accounting for eccentricity, as observed in perturbed orbits, the constellation's evolution deviates from the ideal equilateral triangle, exhibiting periodic variations.

The close-to-circular orbits, as inspired by Eqs.~(\ref{eq:e_0value}) and (\ref{eq:NominalOrbit_con}) for Keplerian orbits, are currently employed in TianQin orbit studies (see, e.g., \cite{Hu2018,Ye2019,Zhou2021,Zhang2021,Luo2022,Yao2023,Jia2023,Zhang2023}).   It is worth noting that, to obtain the nominal equilateral triangle configuration, there is another option: elliptical frozen orbits. From Eqs.~(\ref{eq:L_ij_e1})-(\ref{eq:alpha_k_e1}), if $ e_{1} = e_{2} = e_{3} \equiv  e_{\text{o}}  \neq 0 $, and $ M_{1} = M_{2} = M_{3} $, namely $ \omega_{k} =  \omega_{1} + \frac{2\pi }{3}(k-1) $, then $ L_{ij}^{\text{kepl-B}}(t) \equiv  \sqrt{3} a_{\text{o}} - \sqrt{3} a_{\text{o}} e_{\text{o}} \cos E_{1}(t) $, representing an equilateral-triangle constellation with three arm-lengths that vary synchronously. Preliminary numerical simulation results show that the long-term stability of the constellation, based on this design, is not as favorable as that of the close-to-circular orbits. Furthermore, the impact of this design on other aspects of the mission, such as point-ahead angle variations associated with the finite speed of light, requires further assessment. In this paper, we focus on the study of a three-satellite constellation with close-to-circular orbits.


\section{\label{sec:3} Effects of lunisolar perturbations on TianQin constellation}

The TianQin constellation is subject not only to the central gravitational attraction but also to gravitational perturbations. These perturbations can distort the carefully designed equilateral-triangle configuration. To gain a more accurate understanding of the TianQin constellation's variations, it is crucial to account for these gravitational perturbations. 

\begin{figure*}[htb]
\begin{minipage}{0.46\textwidth} 
\flushright
\includegraphics[width=\textwidth]{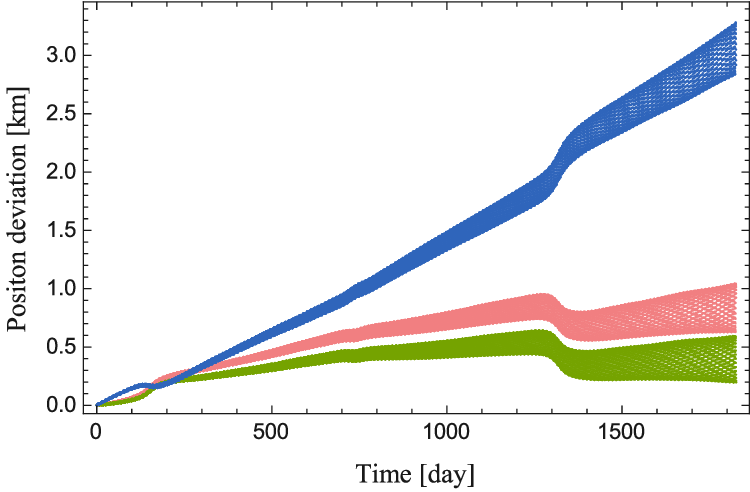}
\end{minipage} 
\hspace{0.2cm}
\begin{minipage}{0.46\textwidth} 
\flushright
\includegraphics[width=0.993\textwidth,height=0.66\textwidth]{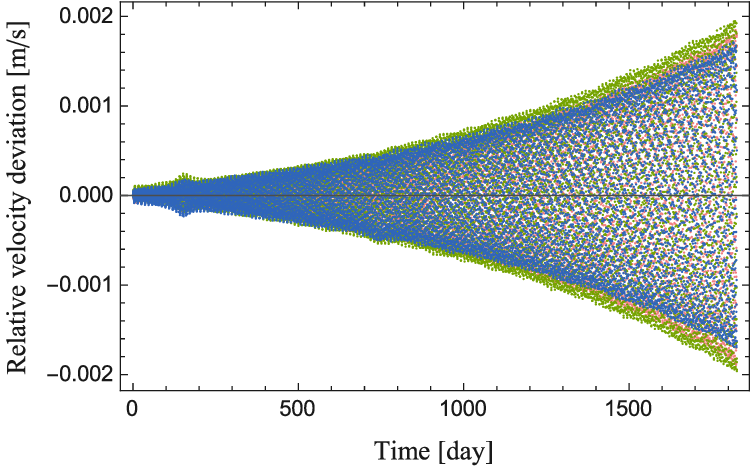}
\end{minipage} 
\begin{minipage}{0.46\textwidth} 
\flushright
\includegraphics[width=0.993\textwidth]{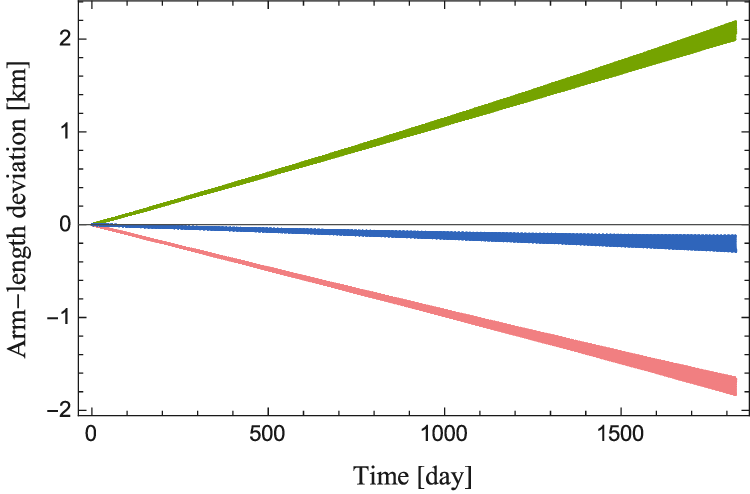}
\end{minipage} 
\hspace{0.2cm}
\begin{minipage}{0.46\textwidth} 
\flushright
\includegraphics[width=\textwidth,height=0.66\textwidth]{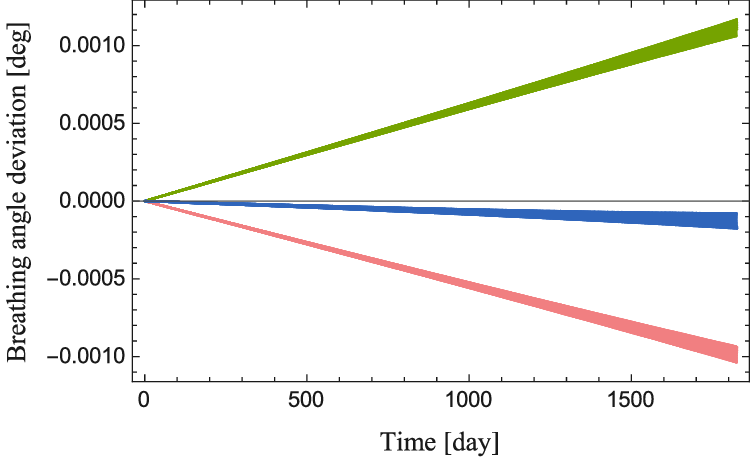}
\end{minipage}
\caption{\label{fig:evolu_simpVSfull} Time evolution of deviations between the simplified (SNM) and high-precision (HPNM) numerical models for satellite positions and three indicators. SNM incorporates the point-mass gravity fields of the Earth, Moon, and Sun, as well as the Earth's $J_{2}$. HPNM additionally considers the higher-degree non-spherical gravity fields of the Earth and the point-mass gravity fields of other planets (for further details, see Appendix \ref{sec:Ana_solution_DerivVerif2}). In these plots, red corresponds to SC1, $v_{23}$, $L_{23}$, or $\alpha_{1}$; green represents SC2, $v_{31}$, $L_{31}$, or $\alpha_{2}$; and blue indicates SC3, $v_{12}$, $L_{12}$, or $\alpha_{3}$. The initial orbital elements used are the same as those presented in Table \ref{tab:orbits_error}.
} 
\end{figure*}

The primary perturbations originate from the Moon and the Sun, with magnitudes of approximately $4 \times 10^{-4}$ and $2 \times 10^{-4}$, respectively \cite{Capderou2014}. In this section, we collectively address the point mass effects of these two perturbing bodies. Furthermore, we also incorporate the secular perturbation arising from the third most significant perturbation, Earth's $ J_{2}  $ perturbation, which has a magnitude of $6 \times 10^{-6}$. Other perturbations, e.g., the higher-degree non-spherical gravity fields of the Earth, have a minor impact on satellite positions and constellation stability. As illustrated in Fig.~\ref{fig:evolu_simpVSfull}, these perturbations lead to deviations of approximately 3.3 km in satellite positions, $ \pm 2.2 $ km in arm-lengths, $ \pm 0.0020 $ m/s in relative velocities, and $ \pm 0.0012^\circ$ in breathing angles over a 5-year period. In contrast to Refs. 
\cite{Hu2018,Qiao2023a,Jiao2023,Xu2011,Roscoe2013,Nie2019}, the perturbation solutions developed in this study offer explicit expressions with improved precision, enabling a more precise description of the distinctive $10^{5}$ km-radius orbits relevant to space-based GW detection.

\subsection{Dynamic model}

The gravitational potential $U$ acting on a satellite can be expressed as
\begin{align}
U = U_{0} + \mathcal{R}, \qquad U_{0} = \frac{\mu}{r},
\end{align}
where $U_0$ is the gravitational potential of a pointlike Earth, and $\mathcal{R}$ represents a perturbative potential describing the satellite's perturbed motion. Under the influence of $\mathcal{R}$, the evolution of the satellite's orbital elements is governed by Lagrange's planetary equations \cite{Xu2013},
\begin{align} 
\frac{d a}{d t}=&~\frac{2 }{n a}\frac{\partial \mathcal{R}}{\partial \lambda}, \label{eq:Eq_a}\\
\frac{d i}{d t} =&~ \frac{1}{n a^2 \check{e} \sin i}\left [ \cos i  \left(\xi  \frac{\partial \mathcal{R}}{\partial \eta } -\eta  \frac{\partial \mathcal{R}}{\partial \xi } +\frac{\partial \mathcal{R}}{\partial \lambda }\right) \right. \nonumber  \\
& \left. -\frac{\partial \mathcal{R}}{\partial \Omega } \right ],\\
\frac{d \Omega}{d t}=&~\frac{1}{n a^2 \check{e} \sin i}\frac{\partial \mathcal{R}}{\partial i},\\
\frac{d \xi}{d t}=&-\frac{\check{e}}{n a^2}\frac{\partial \mathcal{R}}{\partial \eta }-\xi \frac{\check{e}}{n a^2 (1 + \check{e})}\frac{\partial \mathcal{R}}{\partial \lambda }+\eta \cos i \frac{d \Omega}{d t},\\
\frac{d \eta}{d t}=&~ \frac{\check{e}}{n a^2}\frac{\partial \mathcal{R}}{\partial \xi }- \eta\frac{\check{e}}{n a^2 (1 + \check{e})}\frac{\partial \mathcal{R}}{\partial \lambda }-\xi \cos i \frac{d \Omega}{d t},\\
\frac{d \lambda }{d t}=&~ n-\frac{2}{n a}\frac{\partial \mathcal{R}}{\partial a} +\frac{\check{e}}{n a^2 (1 + \check{e})}\left ( \xi \frac{\partial \mathcal{R}}{\partial \xi }+ \eta \frac{\partial \mathcal{R}}{\partial \eta } \right )  \nonumber \\
& -\cos i \frac{d \Omega}{d t},\label{eq:Eq_lambda}
\end{align}
where $ \check{e}:= \sqrt{1-e^2} $, and new variables $\xi$ and $\eta$ are introduced, 
\begin{align}
\xi := e\cos \omega, \qquad \eta :=  e\sin \omega,\label{eq:xieta_e}
\end{align}
to avoid the singularity at $e = 0$. When $\mathcal{R} = 0$, the solutions to Eqs.~(\ref{eq:Eq_a})-(\ref{eq:Eq_lambda}) revert to the Keplerian case discussed in Sec. \ref{sec:2.1}.

For TianQin orbits with an orbital radius of $ 10^{5} $ km, the perturbative potential $ \mathcal{R} $ predominantly encompasses the perturbation effects arising from the Sun, Moon, and Earth's $ J_2 $ term, as expressed in the following expressions:
\begin{align}
\mathcal{R} =  \mathcal{R}_{\text{s}} + \mathcal{R}_{\text{m}} +  \mathcal{R}_{J_{2}} \label{eq:R_smJ2} 
\end{align}
with \cite{Capderou2014}
\begin{align}
& \mathcal{R}_{\text{s}} =  \frac{ \mu _{2} r^{2} }{r_{2}^{3} }  \frac{3 \cos ^{2} \psi_{2} - 1}{2},\label{eq:R_s} \\
& \mathcal{R}_{\text{m}} = \frac{ \mu _{3} }{r_{3}}\sum_{N=2}^{ \mathcal{N} }\left (\frac{r}{r_{3}} \right )^{N} P_{N}(\cos \psi_{3} ),\label{eq:R_m} \\
& \mathcal{R}_{\! J_2} = -  \frac{ \mu R_{\text{e}}^{2}}{r^{3}}  J_2 \frac{3 \sin^{2} \varphi - 1}{2},\label{eq:R_J2}
\end{align}
where $  \mu _{2} = G M_{\text{s}} $ and $  \mu _{3} = G M_{\text{m}} $ are the gravitational constants of the Sun and the Moon, respectively. $r_2$ and $r_3$ denote the geocentric distances of the Sun and the Moon. Moreover, $ P_{N}(x) $ is the Legendre polynomial of degree $N$, with $\mathcal{N} = 6 $ signifying the truncation degree. The derivation of Eq.~(\ref{eq:R_m}) is presented in Appendix \ref{sec:Ana_solution_DerivVerif0}, suggesting that employing Legendre polynomial expansions is more advantageous than the original square root form (Eq.~(\ref{eq:R_m_2})) for solving Lagrange's equations. Additionally, $ R_{\text{e}} $ stands for the equatorial radius of the Earth, and $J_2$ represents the second zonal harmonic coefficient.
Furthermore, $\psi_2$ is the angular separation of the Sun and the satellite as observed from the Earth's center,
\begin{align}
\cos \psi_{2} = \mathbf{\hat{r}_{2}}\cdot   \mathbf{\hat{r}},\label{eq:cos_psi_2}
\end{align}
where $  \mathbf{\hat{r}_{2}}$ and $ \mathbf{\hat{r}} $ denote the unit position vectors of the Sun and the satellite, respectively,
\begin{align}
\mathbf{\hat{r}_{2}}= \begin{bmatrix}
\cos \text{u}_{2} \\ 
\sin \text{u}_{2} \\ 
0
\end{bmatrix},\quad 
\mathbf{\hat{r}}= R_{z}(-\Omega)R_{x}(-i)R_{z}( -\omega ) 
\begin{bmatrix}
\cos \nu \\ 
\sin \nu \\ 
0
\end{bmatrix},\label{eq:r_unit_vec}
\end{align}
with $\text{u}_2 :=  \Omega_2 + \omega_2 + \nu_2 $ representing the Sun's ecliptic longitude.
Similarly, $\psi_3$ is given by 
\begin{align}
 \cos \psi_{3} = \mathbf{\hat{r}_{3}} \cdot   \mathbf{\hat{r}},\label{eq:cos_psi_3}
\end{align}
with $ \mathbf{\hat{r}_{3}} $ being the Moon's unit position vector,
\begin{align}
\mathbf{\hat{r}_{3}}= R_{z}(-\Omega _{3})R_{x}(-i_{3})R_{z}(- \text{u}_{3}) 
\begin{bmatrix}
1\\ 
0\\
0
\end{bmatrix},\label{eq:r3_unit_vec}
\end{align}
where $\Omega_3$, $i_3$, and $ \text{u}_{3} :=  \omega_{3} + \nu _{3}  $ correspond to the Moon's longitude of ascending node, inclination, and latitude argument, respectively. $\varphi$ of Eq.~(\ref{eq:R_J2}) signifies the satellite's geocentric latitude in the Earth-fixed coordinate system,
\begin{align}
\sin \varphi  =\sin i \sin (\omega + \nu).\label{eq:sin_varphi}
\end{align}
Substituting Eqs.~(\ref{eq:cos_psi_2})-(\ref{eq:sin_varphi}) into (\ref{eq:R_s})-(\ref{eq:R_J2}), $\mathcal{R}$ is formulated as a function of the satellite's orbital elements and those of the Sun and the Moon, $ \mathcal{R}(a,e,i,\Omega,\omega,\nu; \sigma _{23}) $, where $ \sigma _{23} = \{ r_{2},\text{u}_{2},i_{3},\Omega  _{3},r_{3},\text{u}_{3} \} $. Consequently, the singularity-free form of the potential, $ \mathcal{R}(a,\xi,\eta,i,\Omega,\lambda; \sigma _{23}) $, can be obtained. Moreover, it is important to note that Eq.~(\ref{eq:sin_varphi}) is formulated in the equatorial coordinate system. For a unified description of the influence of all three perturbations on satellite orbits, including both solar and lunar perturbations, the transformation to the ecliptic coordinate system must be taken into account (see Appendix \ref{sec:Ana_solution_J2} for more details).

\subsection{Motion of the Sun and Moon}   \label{sec:MS_parameter_simplify}

To solve the Lagrange equations, the coordinates of the Sun and the Moon, relative to $r_2$, $\text{u}_2$, $i_3$, $\Omega_3$, $r_3$, and $\text{u}_3$, are required. While high-precision coordinates are available through numerical integration, such as the JPL ephemerides DE430 \cite{Folkner2014}, they are less suitable for analytical purposes. References like \cite{Liu2000,Montenbruck2001} offer analytical formulas with reduced precision, providing geocentric solar coordinates based on a simplified, unperturbed motion of the Earth around the Sun and expressed using appropriate mean orbital elements. In contrast, the Moon's motion, influenced by strong solar and terrestrial perturbations, is described through linear terms corresponding to its long-term precessing elliptical orbit and numerous trigonometric terms capturing periodic variations. 

Generally, higher precision in these analytical coordinates results in more complex expressions, rendering the analytical solutions of the Lagrange equations challenging. In this study, for a balance between solvability and precision, essential components of these coordinates are retained, and fitting is applied using JPL ephemerides data \cite{JPLdata} from around 2035 to reduce discrepancies in the positions of the Sun and the Moon.

The apparent motion of the Sun around the Earth is approximated as a circular orbit on the ecliptic plane with a one-sidereal-year period, 
\begin{align}
& r_2 (t) \simeq \overline{r}_{2} , \\ 
& \text{u}_{2}(t) \simeq u_2 (t) = n_{2} \, t + u_{2_{0}},
\end{align}
where $ \overline{r}_{2} $ is the mean Sun-Earth distance, and $u_2$ represents the mean longitude of the Sun. Moreover, $ n_{2} =  2\pi/(\text{one sidereal year}) $ is the mean motion, and $u_{2_{0}}$ denotes the initial phase of the Sun's orbit. The specific parameter values can be found in Table \ref{table:settings}. The Moon's orbit is considered as an inclined and elliptical precessing orbit,
\begin{align}
& i_{3}(t) \simeq  \overline{i}_{3},\\
& \Omega_{3}(t) \simeq  \overline{\Omega} _{3}(t), \\ 
& r_3(t) \simeq \overline{r}_{3} + r_3^{A}\cos \overline{M}_{3}(t),\label{eq:Delta_u3_M3}\\ &
 \text{u}_{3}(t)  \simeq  u_{3}(t) + u_3^{A}\sin \overline{M}_{3}(t),\label{eq:Delta_u3_u3} 
\end{align}
with
\begin{align}
& \overline{\Omega} _{3}(t) =  n_{\Omega_{3}} t + \Omega_{3_{0}}, \\ 
& \overline{M}_{3}(t) = n_{\! M_3} t + M_{3_{0}}, \\ 
& u_{3}(t) = n_{3} \, t + u_{3_{0}}, 
\end{align}
where $ \overline{i}_{3} $ is the mean inclination of the Moon's orbit, $ \overline{\Omega} _{3} $ represents the secular variation in the Moon's longitude of ascending node, $ \overline{r}_{3} $ denotes the mean Earth-Moon distance, and $ u_{3} $ signifies the secular variation in the argument of latitude. Equations (\ref{eq:Delta_u3_M3}) and (\ref{eq:Delta_u3_u3}) include trigonometric corrections, related to the Moon's mean anomaly $\overline{M}_{3}$, aimed at more accurately describing the Moon's motion in the radial and transverse directions. The periods of variation for $\overline{\Omega}_ {3}$, $\overline{M}_{3}$, and $u_{3}$ are approximately 18.6 years, 27.55 days (anomalistic month), and 27.21 days (draconic month), respectively.

\begin{table}[htb]
\caption{\label{table:settings} Parameter settings for the motion of the Sun and Moon. The subscript ``0" indicates values taken at the epoch 1 January, 2034, 00:00:00 UTC.}
\begin{ruledtabular}
\begin{tabular}{llcc}
Symbols &  \multicolumn{1}{c}{Parameters}  &  Values \\ 
\hline
$ \overline{r}_{2}  $  & Mean Sun-Earth distance & $ 1.496\,191 \times 10^{8} \, \text{km} $  \\ 
$ n_{2} $  & Mean motion of the Earth  & $ 2\pi/365.2564 \, \text{days} $ \\ 
$ u_{2_{0}} $  & Sun initial phase & $ 280.251^\circ $ \\
$ \overline{i}_{3} $  & Mean lunar orbit inclination & $ 5.162^\circ $ \\ 
$ n_{\Omega_{3}} $  & Rate of change of $ \overline{\Omega}_{3}$ &  $ 2\pi/18.6 \, \text{years}  $ \\ 
$ \Omega_{3_{0}} $  & Initial phase of $ \overline{\Omega}_{3}$ & $ 186.988^\circ $ \\
$ \overline{r}_{3} $  & Mean Earth-Moon distance & $ 384\,151 \, \text{km} $ \\
$ r_3^{A} $  & Amplitude of correction term & $ -20\,905 \, \text{km} $ \\
$ n_{\! M_3} $  & Rate of change of $ \overline{M}_{3}$
 &  $ 2\pi/27.55 \, \text{days} $ \\ 
$ M_{3_{0}} $  & Initial phase of $ \overline{M}_{3}$ & $ 22.578^\circ $ \\
$ n_{3} $  & Rate of change of $ u_{3}$ & $ 2\pi/27.21 \, \text{days} $  \\ 
$ u_{3_{0}} $  & Initial phase of $ u_{3}$ & $ 221.970^\circ $ \\
$ u_3^{A} $  & Amplitude of correction term  & $ 6.289^\circ $ \\
\end{tabular}
\end{ruledtabular}
\end{table}

\subsection{Lunisolar perturbations on the TianQin satellites}

Let $\sigma = \begin{bmatrix} 
a & i & \Omega & \xi & \eta & \lambda
\end{bmatrix}^{T}$; then, the Lagrange perturbation equations (\ref{eq:Eq_a})-(\ref{eq:Eq_lambda}) can be reformulated as
\begin{align}
\frac{d \sigma }{d t} = f_{0}(a)+f_{1}(\sigma,t,\varepsilon),\label{eq:perturbation}
\end{align}
where the functions $f_{0}$ and $f_{1}$ are both 6-dimensional vector functions,
\begin{align}
& f_{0}(a)= \delta_{\iota} n,\qquad  \delta_{\iota} = \begin{bmatrix}
0 &  0 & 0  &  0 &  0  & 1
\end{bmatrix}^{T},\label{eq:delta_01} \\
& ( f_{1})_{\zeta}=\mathcal{O}(\varepsilon ),\qquad \zeta =1,2,\cdots ,6,
\end{align}
and $\varepsilon \ll 1$ is a small parameter. Since the perturbing forces are significantly weaker than the Earth's central gravitational attraction, the solution of Eq.~(\ref{eq:perturbation}) is assumed to be
\begin{align}
\sigma (t)  = \sigma^{(0)} (t) + \sigma^{(1)} (t).\label{eq:sol_form1}
\end{align}
Here, $\sigma^{(0)} (t)$ represents the unperturbed Keplerian orbit (as shown in Eq.~(\ref{eq:sol_unperturbed_sec2})),
\begin{align}
\sigma^{(0)} (t) = \sigma_{0} +\delta_{\iota} n(t-t_{0}),\label{eq:sol_unperturbed}
\end{align}
and $\sigma^{(1)}(t)$ is the perturbation solution,
\begin{align}
\sigma^{(1)} (t) = \bigl (\sigma _{c}(t)  + \sigma _{l[c]}(t) + \sigma _{l[l]}(t)  + \sigma _{s}(t) \bigr ) \!\large{|}_{t_{0}}^{t}.\label{eq:sol_perturbed}
\end{align}
In Eq.~(\ref{eq:sol_perturbed}), $\sigma^{(1)}(t)$ is decomposed into four parts, distinguished by the unique time scales of orbital variations induced by perturbations: the secular term $\sigma_{c}$, special long-period term $\sigma_{l[c]}$, general long-period term $\sigma_{l[l]}$, and short-period term $\sigma_{s}$. $\sigma_{c}$ signifies the linear change over time, while $\sigma_{l[c]}$, $\sigma_{l[l]}$, and $\sigma_{s}$ are associated with periodic variations. These variations are linked to, for instance, $\Omega_{3}$ with an 18.6-year period, $u_{3}$ with a 27.21-day period, and $\lambda$ with a 3.64-day period. The explicit expression for $\sigma^{(1)} (t)$ can be derived using perturbation methods to solve Eq.~(\ref{eq:perturbation}). 

\begin{table}[ht]
  \caption{The components of the analytical solution $ \sigma(t) $ for the TianQin orbit describing the perturbing effects from the Sun, Moon, and Earth's $ J_{2} $. The symbol ``$-$" denotes that terms do not appear separately due to the joint effects of solar and lunar perturbations. ``$\sim$" indicates neglected contributions, considering that the $ J_{2} $ perturbation induces a negligible eccentricity variation of $ 10^{-6} $. Terms like $ \sigma _{J_{2}}^{s} $ are also ignored, and $ \sigma _{J_{2}}^{l[l]} = \sigma _{J_{2}}^{l[c]} =0 $. Explicit expressions for each component are detailed in Appendix \ref{sec:Ana_solution_expression}.}\label{table:solution_all_use}
\begin{ruledtabular}
\begin{tabular}{cccccccccccccc}
      \multirow{2}*{ $ \sigma(t) $ }      &  \multirow{2}*{ $ \sigma _{0} + \delta_{\iota}  \overline{n}\,t $  } &  \multicolumn{3}{c}{$ \sigma _{c}(t) $}  &  \multicolumn{2}{c}{$ \sigma _{l[c]}(t) $}  &  \multicolumn{2}{c}{$ \sigma _{l[l]}(t) $}  &  \multicolumn{2}{c}{$ \sigma _{s}(t) $}     \\      
    \cmidrule(r){3-5}  \cmidrule(r){6-7}  \cmidrule(r){8-9}  \cmidrule(r){10-11}
       ~    &    ~    &  s   &  m  & \scriptsize{$ J_{2} $}  &  s   &  m  &  s   &  m   &  s   &  m   \\
        \hline                
$ a(t) $    & $ a_{0} $  &  0 &   0 &  0  &  0 &   0 &  0 &   0 & $ a_{\text{s}}^{s} $ &   $ a_{\text{m}}^{s} $  \\
 $ i(t) $    & $ i_{0} $  &  0 &   0 &  $ i _{J_{2}}^{c}  $  &  0  &   $ i_{\text{m}}^{l[c]} $  &  $ i_{\text{s}}^{l[l]} $ &   $ i_{\text{m}}^{l[l]} $ &  $ i_{\text{s}}^{s} $ &   $ i_{\text{m}}^{s} $  \\
 $ \xi(t) $    & $ \xi_{0} $  &  $-$ &   $ \xi_{c} $ &  $\sim$   &  $-$ &   $ \xi_{l[c]} $   &  0 &   $ \xi_{\text{m}}^{l[l]} $ &  $ \xi_{\text{s}}^{s} $ &   $ \xi_{\text{m}}^{s} $   \\ 
 $ \eta(t) $    & $ \eta_{0} $  &  $-$ &   $ \eta_{c} $ &  $\sim$  &  $-$ &   $ \eta_{l[c]} $    &  0 &   $ \eta_{\text{m}}^{l[l]} $  &   $ \eta_{\text{s}}^{s} $ &   $ \eta_{\text{m}}^{s} $   \\  
  $ \Omega(t) $    & $ \Omega_{0} $ &  $ \Omega_{\text{s}}^{c} $ &   $ \Omega_{\text{m}}^{c} $ &  $ \Omega _{J_{2}}^{c}  $  &  0 &   $ \Omega_{\text{m}}^{l[c]} $ & $ \Omega_{\text{s}}^{l[l]} $ &   $ \Omega_{\text{m}}^{l[l]} $ &  $ \Omega_{\text{s}}^{s} $ &   $ \Omega_{\text{m}}^{s} $  \\
$ \lambda(t)  $ & $ \lambda _{0} +\overline{n}\,t $ &  $ \lambda _{\text{s}}^{c} $ &   $ \lambda _{\text{m}}^{c} $ &  $ \lambda  _{J_{2}}^{c}  $  &  0 &   $ \lambda _{\text{m}}^{l[c]} $ & $ \lambda _{\text{s}}^{l[l]} $ & $ \lambda _{\text{m}}^{l[l]} $  &  $ \lambda _{\text{s}}^{s} $ &   $ \lambda _{\text{m}}^{s} $  \\
    \end{tabular} 
    \end{ruledtabular}  
\end{table}

To enhance the precision of the analytical solution, we employ a perturbation method known as the mean element method \cite{Kozai1973,Liu2000}, which uses the mean orbital elements $ \overline{\sigma} (t) $ as a reference solution (defined in Eq.~(\ref{eq:sigmaPJt_NiPJ})), rather than the Keplerian orbit $ \sigma^{(0)} (t) $. Additionally, for simplicity, only terms up to the first order of eccentricity in the solution are considered. For more derivation details, one can refer to Appendix \ref{sec:Ana_solution_DerivVerif12}. The components of the perturbation solution are outlined in Table \ref{table:solution_all_use}, with the average value $ \overline{n} = \sqrt{\frac{\mu}{\overline{a}^{3}}} $ replacing $ n $ in $ \sigma^{(0)} (t) $. Detailed expressions for the terms in Table \ref{table:solution_all_use} can be found in Appendix \ref{sec:Ana_solution_expression}. Moreover, the effectiveness of the analytical solution $ \sigma (t) $ is evaluated through a comparison with high-fidelity numerical orbit simulations (see Appendix \ref{sec:Ana_solution_DerivVerif2}). For the TianQin orbit, the 5-year average deviation in position is approximately 87 km.

Table \ref{table:solution_all_use} illustrates the effects of gravitational perturbations on the orbital elements of the TianQin satellite. Variation in $ a $ are solely induced by short-period perturbation. However, the other five orbital elements are also influenced by both secular and long-period perturbations, particularly the secular one, leading to cumulative change. In the case of $i$, its secular variation are not due to lunisolar perturbations (which would occur when considering second-order eccentricity \cite{Prado2003}), but instead result from $ J_{2}  $ perturbation, tied to a coordinate transformation involving the obliquity $\epsilon$ (see Eq.~(\ref{eq:i_J2_c})). As for $\Omega$, $\lambda$, $\xi$, and $\eta$, their secular variations are predominantly driven by lunar and solar perturbations. 

The two elements, $\Omega$ and $i$, determine the orientation of the orbital plane. As indicated in Eqs.~(\ref{eq:sol_c_sun_raan}) and (\ref{eq:sol_c_moon_raan}), the secular variation of $\Omega$ is dependent on the satellite's mean semimajor axis $\overline{a}$ and mean inclination $\overline{i}$, which implies that $\Omega$ experiences negligible precession when $i\sim 90^{\circ}$. Similarly, the secular variation of $i$ for the TianQin satellite is also minimal ($< 0.1^\circ $ in five years). As a result, the orientation of the TianQin detector plane remains nearly constant, changing by less than $ 2.6^{\circ} $ over five years. This is in stark contrast to LISA orbits \cite{LISA2017}, where the plane undergoes a full 360-degree rotation annually.

For the periodic variations, their periods are linked to the motions of the satellite, the Moon, and the Sun. Especially, for the short-period variation $ \sigma _{s}(t) $, the arguments of trigonometric terms take the form of $\kappa \,\lambda + p\,u_{3} + q\,\theta_{3}$ or $\kappa\, \lambda + p \, U_{2}$ (see, e.g., Eqs.~(\ref{eq:as_m}) and (\ref{eq:sma_s_sun})), which indicates that orbital variations occur at multiples of the satellite's orbital frequency and are modulated by the motions of the Moon and the Sun. This insight aids in understanding the perturbing effects of the Moon's and the Sun's gravitational fields on the TianQin inter-satellite range acceleration noise (cf.~Fig.~3 in Ref.~\cite{Zhang2021}). 

Moreover, $ \sigma_{s}(t) $ is correlated with the orbit phase $ \lambda_{k}(t) $ of SC$k$, with a 120-degree phase difference among the three satellites, indicating that short-period perturbation will influence the relative motion between satellites. However, it can be demonstrated that the other three components of Eq.~(\ref{eq:sol_perturbed}) may have little impact on the relative motion. Furthermore, an ideal equilateral-triangle constellation requires zero eccentricity, which is unlikely to hold, as shown by the perturbation solutions $ \xi_{k}^{(1)}(t) $ and $ \eta_{k}^{(1)}(t)$. Consequently, the presence of these components underscores the potential to significantly disturb the nominal TianQin triangle constellation.

\subsection{\label{sec:3.4} Perturbed motion of the TianQin constellation}

Equations (\ref{eq:sol_form1})-(\ref{eq:sol_perturbed}) describe the variations in orbital elements for SC$ k $ ($  k$ = 1, 2, 3) under the influence of lunisolar perturbations and Earth's $ J_{2}  $ perturbation in the geocentric ecliptic coordinate system,
\begin{align}
\sigma_{k}(t) = &~ \sigma _{0k} + \delta_{\iota} \, \overline{n}_{k}(t-t_{0}) \nonumber\\
& + \bigl (\sigma_{k}^{c}(t)  + \sigma_{k}^{l[c]}(t) + \sigma_{k}^{l[l]}(t)  + \sigma_{k}^{s}(t) \bigr ) \!\large{|}_{t_{0}}^{t},\label{eq:sigma_msJ2}
\end{align}
with their explicit expressions detailed in Table \ref{table:solution_all_use} and Appendix \ref{sec:Ana_solution_expression}. Substituting Eq.~(\ref{eq:sigma_msJ2}) into Eq.~(\ref{eq:z_k}) and using Eq.~(\ref{eq:xieta_e}), the position vector $ \mathbf{r}_{k} $ and velocity vector $ \mathbf{\dot{r}}_{k} $ can be obtained. Then, employing the definitions in Eqs.~(\ref{eq:L_ij})-(\ref{eq:alpha_k}), one can derive analytical expressions for the constellation's three kinematic indicators, $L_{ij}(t)$, $v_{ij}(t)$, and $\alpha_{k}(t) $. The time evolution of these three quantities is illustrated in Fig.~\ref{fig:evolu_anaVSnum} for a set of simulated TianQin orbits, comparing both analytical and numerical models (see Appendix \ref{sec:Ana_solution_DerivVerif3} for more details).

Variations in the triangular constellation can be decomposed into two parts,
\begin{align}
 L_{ij}(t)=&~ L_{\text{o}} + \delta  L_{ij}(t),\label{eq:Lij_decompose} \\
 v_{ij}(t)=&~ v_{\text{o}} + \delta  v_{ij}(t), \\ 
\alpha_{k}(t)=&~ \alpha_{\text{o}} + \delta  \alpha_{k}(t),
\end{align}
where $L_{\text{o}} = \sqrt{3} \times 10^5$ km, $ v_{\text{o}}=0 $ m/s, and $\alpha_{\text{o}} = \frac{\pi}{3}$ represent the desired equilateral-triangle configuration. Conversely, $\delta L_{ij}(t)$, $\delta v_{ij}(t)$, and $\delta \alpha_{k}(t)$ signify distortions from the ideal configuration. As previously mentioned, the magnitude of these distortions significantly impacts GW detection missions, including TDI data processing and the design of instruments such as phase meters and telescopes. It is crucial to minimize these distortions as much as possible.

These distortions are all zero when the three satellites are solely influenced by the Earth's point mass and satisfy the conditions (\ref{eq:e_0value}) and (\ref{eq:NominalOrbit_con}). However, these conditions no longer hold when accounting for gravitational perturbations, which result in variations in $ e_{k}(t) $ and the inclusion of non-synchronous short-period terms $ \sigma_{k}^{s}(t) $ in $ \sigma_{k}(t) $. To achieve an equilateral triangle, $ \sigma_{k} (t) $ in Eqs.~(\ref{eq:e_0value}) and 
(\ref{eq:NominalOrbit_con}), instead, can be substituted with $ \sigma_{k}^{\text{ref}}(t) $:
\begin{align}
\left\{
\begin{array}{l}
 e_{1}^{\text{ref}}(t) = e_{2}^{\text{ref}}(t) = e_{3}^{\text{ref}}(t) \equiv e_{\text{o}} = 0,\smallskip\\
 a_{1}^{\text{ref}}(t) = a_{2}^{\text{ref}}(t) = a_{3}^{\text{ref}}(t) \equiv  a_{\text{o}} =\frac{L_{\text{o}}}{\sqrt{3}}, \smallskip\\
i_{1}^{\text{ref}}(t) = i_{2}^{\text{ref}}(t) = i_{3}^{\text{ref}}(t) \equiv i_{\text{o}}(t),\smallskip\\
\Omega_{1}^{\text{ref}}(t) = \Omega_{2}^{\text{ref}}(t) = \Omega_{3}^{\text{ref}}(t) \equiv \Omega_{\text{o}}(t),\smallskip\\
\lambda_{k}^{\text{ref}}(t) \equiv  \frac{2\pi }{3}(k-1) + \lambda_{\text{o}}(t),
\end{array} 
\right.\label{eq:nomOrb_con_real}
\end{align}
where
\begin{equation}
 \sigma_{k}^{\text{ref}}(t) := \left\{
\begin{array}{ll}
0, & \sigma = e, \smallskip\\
\sigma_{k}(t) - \sigma_{k}^{s}(t), & \sigma \in \{a,i, \Omega, \lambda \}, \label{eq:nomOrb_con_real_ref}
\end{array}
\right.
\end{equation}
and $ \sigma_{\text{o}}(t) $ serves as the reference orbit for the synchronous motion of the three satellites. Utilizing Eqs.~(\ref{eq:sigma_msJ2}), (\ref{eq:nomOrb_con_real}), and (\ref{eq:nomOrb_con_real_ref}), the form of $ \sigma_{\text{o}}(t) $ for $ \sigma \in \{a,i, \Omega, \lambda \} $ is
\begin{align} 
\sigma_{\text{o}}(t) = \overline{\sigma}_{0\text{o}} + \delta_{\iota} \overline{n}_{\text{o}}(t-t_{0}) + \Delta \sigma_{\text{o}}^{c}(t)  + \Delta \sigma_{\text{o}}^{l[c]}(t) +  \sigma_{\text{o}}^{l[l]}(t),\label{eq:nomOrb_real}
\end{align} 
with $ \overline{\sigma}_{0\text{o}} := \sigma _{0k} - [\sigma_{\text{o}}^{l[l]}(t_{0})+  \sigma_{k}^{s}(t_{0}) ] - \delta _{\iota} \frac{2\pi }{3}(k-1) = \text{const}.$, $ \Delta \sigma_{\text{o}}^{c}(t) := \sigma_{\text{o}}^{c}(t) -\sigma_{\text{o}}^{c}(t_{0}) $, and $ \Delta \sigma_{\text{o}}^{l[c]}(t) := \sigma_{\text{o}}^{l[c]}(t) - \sigma_{\text{o}}^{l[c]}(t_{0}) $. In other words, the three satellites move along the same virtual circular orbit $ \sigma_{\text{o}}(t) $ with secular and long-period variations, while maintaining a 120-degree phase difference, forming a rotating, precessing equilateral-triangle constellation.

Correspondingly, $ \delta L_{ij}(t) $, $ \delta v_{ij}(t) $, and $ \delta \alpha_{k}(t) $ result from discrepancies $ \delta \sigma_{k}(t) $:
\begin{align} 
\delta \sigma_{k}(t):= \sigma_{k}(t)- \sigma_{\text{o}k}(t), \label{eq:orbit_same_k-vs-nomOrb_real}
\end{align}
between the real orbits $ \sigma_{k}(t) $ and the reference orbits $ \sigma_{\text{o}k}(t) = \sigma_{\text{o}}(t)+ \delta _{\iota} \frac{2\pi }{3}(k-1) $.
By expanding $ L_{ij}(t) $, $ v_{ij}(t) $, and $ \alpha_k (t) $ into a Taylor series along $ \sigma_{\text{o}k}(t) $, the triangle distortions caused by $ \delta \sigma_{k} (t) $ can be obtained. For the arm-length distortion, we have
\begin{widetext}
\begin{align}
 \delta L_{ij}(t) = & ~ L_{ij}(t) - \sqrt{3}a_{\text{o}} \nonumber \\ 
= & ~ \frac{\sqrt{3}}{2} [{\delta a}_i(t)+{\delta a}_j(t) ] +\frac{1}{2} a_{\text{o}} \delta \lambda _{ji}(t) +\frac{1}{2} a_{\text{o}} \cos i_{\text{o}}(t) \, \delta \Omega _{{ji}}(t)  \nonumber \\ 
& +\frac{\sqrt{7}}{2}  a_{\text{o}} [\sin M_{\text{o}j}^{-}(t) \, {\delta e}_j(t)-\sin  M_{\text{o}i}^{+}(t) \, {\delta e}_i(t)] +  \mathcal{O}(\delta \sigma(t)^{2}), \label{eq:delta_Lij}
\end{align}
where 
\begin{align} 
\delta \sigma_{ji} (t) :=  \delta \sigma_{j} (t) -\delta \sigma_{i} (t),\label{eq:delta_orb}
\end{align}
\begin{align} 
M_{\text{o}k}^{\pm }(t) := M_{\text{o}k}(t) \pm \beta,\qquad M_{\text{o}k}(t) = \lambda_{\text{o}k}(t) - \omega_{k}(t),
\end{align}
and the indices $ i $, $ j $, and $ k $ follow a cyclic permutation ($ i, \,j,\, k = 1 \to 2 \to 3 \to 1$). It can be seen from Eq.~(\ref{eq:delta_Lij}) that, up to $ (\delta \sigma)^{1} $ order, $ \delta L_{ij}(t) $ is unaffected by inclination deviations $ \delta i_{i} $ and $ \delta i_{j} $. Additionally, deviations in $ \Omega $ have minimal influence on $ \delta L_{ij}(t) $ due to the approximately $ 90 ^{\circ} $ inclinations of TianQin orbits, which render the constellation stability insensitive to orbital plane deviations. Combining Eqs.~(\ref{eq:delta_orb}), (\ref{eq:orbit_same_k-vs-nomOrb_real}), (\ref{eq:sigma_msJ2}), and (\ref{eq:nomOrb_real}), and defining $ i_{\text{o}}(t) = \overline{i}_{\text{o}} + i^{\epsilon}_{\text{o}}(t)   $, where $ \overline{i}_{\text{o}} $ is the average, the right-hand side of Eq.~(\ref{eq:delta_Lij}) can be categorized into distinct types:
\begin{align} 
 \delta L_{ij}(t) = &~ \delta L_{ij}^{\text{drift}}(t) + \delta L_{ij}^{\text{bias}} + \delta L_{ij}^{\text{fluc}}(t)  + \mathcal{O}(i^{\epsilon }_{\text{o}}(t)^{1}) \delta \Omega _{{ji}}(t) +  \mathcal{O}(\delta \sigma(t)^{2}), \label{eq:delta_Lij_3type}
\end{align}
with
\begin{align}
 \delta L_{ij}^{\text{drift}}(t) =&~ \frac{1}{2} a_{\text{o}} \delta \overline{n}_{ji}(t-t_{0}) + \frac{1}{2} a_{\text{o}}[\delta_{\!\Delta} \lambda_{ji}^{c}(t) + \delta_{\!\Delta} \lambda_{ji}^{l[c]}(t) ]+ \frac{1}{2} a_{\text{o}} \cos \overline{i}_{\text{o}} \, [ \delta_{\!\Delta} \Omega_{ji}^{c}(t) + 
\delta_{\!\Delta} \Omega_{ji}^{l[c]}(t)],\\
 \delta L_{ij}^{\text{bias}} =&~  \frac{\sqrt{3}}{2} ( \delta \overline{a}_{0i} + \delta \overline{a}_{0j} ) + \frac{1}{2} a_{\text{o}} \delta \overline{\lambda} _{0{ji}} + \frac{1}{2} a_{\text{o}} \cos \overline{i}_{\text{o}} \, \delta \overline{\Omega} _{0{ji}} , \\
\delta L_{ij}^{\text{fluc}}(t) =&~   \frac{\sqrt{7}}{2}  a_{\text{o}} [\sin M_{\text{o}j}^{-}(t) \, e_j(t)-\sin  M_{\text{o}i}^{+}(t) \, e_i(t)]  + \frac{\sqrt{3}}{2}[a^{s}_i(t) +  a^{s}_j(t)] \nonumber \\ 
&  +  \frac{1}{2} a_{\text{o}} [   \delta \lambda_{ji}^{l[l]}(t)+ \delta \lambda ^{s}_{ji}(t) ]  +\frac{1}{2} a_{\text{o}} \cos \overline{i}_{\text{o}} \, [   \delta \Omega_{ji}^{l[l]}(t)+ \delta \Omega^{s}_{ji}(t) ],
\end{align}
where $ \delta \overline{n}_{ji} = \overline{n}_{j} - \overline{n}_{i} $,
\begin{align}
\delta_{\!\Delta}  \sigma_{ij}(t) := \delta \sigma_{ij}(t) - \delta \sigma_{ij}(t_{0}), \label{eq:delta_Delta}
\end{align} 
$ \delta \overline{a} _{0{i}} = \overline{a} _{0{i}} -a _{\text{o}} $, $ \delta \overline{\lambda} _{0{ji}} = \overline{\lambda} _{0{j}} -\overline{\lambda} _{0{i}} - \frac{2\pi }{3}(j-i) $,  and so on. 
Equation (\ref{eq:delta_Lij_3type}) illustrates that arm-length distortion manifests in three possible types: linear drift $ \delta L_{ij}^{\text{drift}}(t) $, constant bias $ \delta L_{ij}^{\text{bias}} $, and periodic fluctuation $ \delta L_{ij}^{\text{fluc}}(t) $. Specifically, (1) $ \delta L_{ij}^{\text{drift}}(t) $ consists of five components of inter-satellite deviations, including $ \delta \overline{n}_{ji} $, $ \delta_{\!\Delta} \lambda_{ji}^{c} $, $ \delta_{\!\Delta} \Omega_{ji}^{c} $, $ \delta_{\!\Delta} \lambda_{ji}^{l[c]} $, and $ \delta_{\!\Delta} \Omega_{ji}^{l[c]} $; (2) $ \delta L_{ij}^{\text{bias}} $ comprises initial mean deviations, $ \delta \overline{a}_{0i} $, $ \delta \overline{a}_{0j} $, $ \delta \overline{\lambda} _{0{ji}} $, and $ \delta \overline{\Omega}_{0ji} $; and (3) $ \delta L_{ij}^{\text{fluc}}(t) $ is linked to eccentricity variations $ e_i (t) $ and $ e_j (t) $ and short-period variations in semi-major axis $ a^{s}_i (t) $ and $ a^{s}_j (t) $, along with inter-satellite periodic deviations in $ \lambda $ and $ \Omega $. Among these types, the drift, which progressively increases over time, emerges as the predominant factor affecting the stability of the constellation.

Regarding the relative velocity, one has
\begin{align}
\delta v_{ij}(t) = & ~  v_{ij}(t) - 0 \nonumber \\ 
= &  -\frac{3}{4}\frac{v_{\text{o}}}{ a_{\text{o}}} [{\delta a}_j(t) - {\delta a}_i(t) ] +\frac{\sqrt{7}}{2} v_{\text{o}} [\cos M_{\text{o}j}^{-}(t) \, {\delta e}_j(t)-\cos  M_{\text{o}i}^{+}(t) \, {\delta e}_i(t)] +  \mathcal{O}(\delta \sigma(t)^{2}),\label{eq:delta_vij} \\
=&~ \delta v_{ij}^{\text{bias}} + \delta v_{ij}^{\text{fluc}}(t) +  \mathcal{O}(\delta \sigma(t)^{2}),\label{eq:delta_Vij_3type}
\end{align}
where $ v_{\text{o}} := \sqrt{\frac{\mu}{a_{\text{o}}}} $, and
\begin{align}
\delta v_{ij}^{\text{bias}} =&  -\frac{3}{4}\frac{v_{\text{o}}}{ a_{\text{o}}}( \delta \overline{a}_{0j} - \delta \overline{a}_{0i} ),\\
\delta v_{ij}^{\text{fluc}}(t) = & -\frac{3}{4}\frac{v_{\text{o}}}{ a_{\text{o}}}[a^{s}_j(t) -  a^{s}_i(t)] + \frac{\sqrt{7}}{2} v_{\text{o}} [\cos M_{\text{o}j}^{-}(t) \, { e}_j(t)-\cos  M_{\text{o}i}^{+}(t) \, { e}_i(t)].
\end{align}
Equation (\ref{eq:delta_Vij_3type}) illustrates that there is little long-term variation in relative velocity, $ \delta v_{ij}^{\text{drift}}(t) \simeq 0 $, consistent with numerical simulation results (cf.~Fig.~10 in Ref.~\cite{Jia2023}). Additionally, the breathing angle within the TianQin triangle experiences three types of distortion akin to those observed in arm-length: 
\begin{align}
 \delta \alpha_k (t) = & ~ \alpha_{k}(t) - \frac{\pi }{3}  \nonumber \\   
 = & ~ \frac{1}{2 \sqrt{3} a_{\text{o}}}[{\delta a}_i(t)+{\delta a}_j(t)-2 {\delta a}_k(t)] +\frac{1}{2} \delta \lambda_{ji}(t) +  \frac{1}{2} \cos i_{\text{o}}(t) \, \delta \Omega _{{ji}} (t) \nonumber \\
& + \frac{ \sqrt{7}}{6}[f_{e}^{k}(t) {\delta e}_k(t) + f_{e}^{i}(t) {\delta e}_i (t) + f_{e}^{j} (t) {\delta e}_j(t)]  +  \mathcal{O}(\delta \sigma(t)^{2}) \label{eq:delta_alphaij} \\ 
=&~  \delta \alpha_k^{\text{drift}}(t) +  \delta \alpha_k^{\text{bias}} +  \delta \alpha_k^{\text{fluc}}(t)   + \mathcal{O}(i^{\epsilon }_{\text{o}}(t)^{1}) \delta \Omega _{{ji}}(t) +  \mathcal{O}(\delta \sigma(t)^{2}), \label{eq:delta_alphaij_3type}
\end{align}
with 
\begin{align}
 \delta \alpha_k^{\text{drift}}(t) = & ~  \frac{1}{2} \delta \overline{n}_{ji}(t-t_{0}) + \frac{1}{2} [\delta_{\!\Delta} \lambda_{ji}^{c}(t) + \delta_{\!\Delta} \lambda_{ji}^{l[c]}(t) ] +\frac{1}{2} \cos \overline{i}_{\text{o}} \, [\delta_{\!\Delta} \Omega_{ji}^{c}(t) + \delta_{\!\Delta} \Omega_{ji}^{l[c]}(t)],\label{eq:delta_alphaij_3type_d} \\ 
 \delta \alpha_k^{\text{bias}} = & ~ \frac{1}{2 \sqrt{3} a_{\text{o}} } ( \delta \overline{a}_{0ik} +  \delta \overline{a}_{0jk} ) + \frac{1}{2} \delta \overline{\lambda} _{0{ji}} + \frac{1}{2} \cos \overline{i}_{\text{o}} \, \delta \overline{\Omega} _{0{ji}},\label{eq:delta_alphaij_3type_b} \\  
 \delta \alpha_k^{\text{fluc}}(t) = & ~ \frac{ \sqrt{7}}{6}[f_{e}^{k}(t) e_k(t) + f_{e}^{i}(t) e_i (t) + f_{e}^{j} (t) e_j(t)]  + \frac{1}{2 \sqrt{3} a_{\text{o}} }[\delta a^{s}_{ik}(t) + \delta a^{s}_{jk}(t)] \nonumber \\ 
&  +  \frac{1}{2} [ \delta \lambda_{ji}^{l[l]}(t)+ \delta \lambda ^{s}_{ji}(t) ]  +\frac{1}{2} \cos \overline{i}_{\text{o}} \, [  \delta \Omega_{ji}^{l[l]}(t)+ \delta \Omega^{s}_{ji}(t) ],\label{eq:delta_alphaij_3type_f} 
\end{align}
where
\begin{align}
f_{e}^{k}(t) = 2\cos  M_{\text{o}k}(t)  \sin  \beta,\qquad f_{e}^{i}(t) = -\sin M_{\text{o}i}^{-}(t)-2 \sin  M_{\text{o}i}^{+}(t) ,\qquad f_{e}^{j}(t) = 2  \sin M_{\text{o}j}^{-}(t)+\sin  M_{\text{o}j}^{+}(t).
\end{align}
Equations (\ref{eq:delta_alphaij_3type_b}) and (\ref{eq:delta_alphaij_3type_f}) show that bias and fluctuation in the breathing angle, as observed from SC$k$, are associated with deviations in $ a $  and $ e $ of all three satellites. However, concerning $ \lambda $ and $   \Omega $, they are exclusively linked to the relative differences between the other two satellites.

The variations in the three types, drift, bias, and fluctuation, all impact the constellation's stability, necessitating optimization. The drift, associated with $ \delta  \overline{n}_{ji} = -\frac{3  n_{\text{o}} }{2 a_{\text{o}}} {\delta  \bar{a}}_{{ji}}  $, can be significantly mitigated by aligning the mean semi-major axes $ \overline{a}_{j}^{\, \text{optim}} = \overline{a}_{i}^{\, \text{optim}} $. More generally, one can see from, e.g., Eqs.~(\ref{eq:delta_alphaij_3type_d})-(\ref{eq:delta_alphaij_3type_f}), that terms within the drift, bias, and long-period fluctuation, are contingent on the mean or initial mean values of parameters $a$, $i$, $\Omega$, and $\lambda$. Additionally, the fluctuation is also correlated with eccentricities, which exhibit secular variations and serve as the primary factor influencing the amplitude of the fluctuation. Thus, the optimization of constellation variations can be achieved by imposing the following conditions:
\begin{align}
\left\{
\begin{array}{l}
\overline{a}_{1}^{\, \text{optim}} = \overline{a}_{2}^{\, \text{optim}}= \overline{a}_{3}^{\, \text{optim}} ,\\
\overline{i}_{1}^{\, \text{optim}} = \overline{i}_{2}^{\, \text{optim}}= \overline{i}_{3}^{\, \text{optim}} ,\\
\overline{\Omega}_{1}^{\, \text{optim}} = \overline{\Omega}_{2}^{\, \text{optim}}= \overline{\Omega}_{3}^{\, \text{optim}},\\
\overline{\lambda}_{k}^{\, \text{optim}} = \frac{2\pi }{3}(k-1) + \overline{\lambda}_{1}^{\, \text{optim}},\\
e_{k}^{\text{optim}}(t) \simeq 0.
\end{array} 
\right 
.\label{eq:optim_con}
\end{align}
Further, the optimized indicators, up to the leading order, are
\begin{align}
 \delta L_{ij}^{\text{optim}}(t) = & ~ \frac{\sqrt{7}}{2}  a_{\text{o}} [\sin M_{\text{o}j}^{-}(t) \, e_j(t)-\sin  M_{\text{o}i}^{+}(t) \, e_i(t)]   + \frac{\sqrt{3}}{2}[a^{s}_i(t) + a^{s}_j(t)] +  \frac{1}{2} a_{\text{o}} \delta \lambda ^{s}_{ji}(t)   +\frac{1}{2} a_{\text{o}} \cos \overline{i}_{\text{o}} \, \delta  \Omega^{s}_{ji}(t), \label{eq:delta_Lij_optim}
\end{align}
\begin{align}
\delta v_{ij}^{\text{optim}}(t) = & ~ \frac{\sqrt{7}}{2} v_{\text{o}} [\cos M_{\text{o}j}^{-}(t) \, { e}_j(t)-\cos  M_{\text{o}i}^{+}(t) \, { e}_i(t)] -\frac{3}{4}\frac{v_{\text{o}}}{ a_{\text{o}}}[a^{s}_j(t) -  a^{s}_i(t)],\label{eq:delta_Vij_optim}
\end{align}
\begin{align}
 \delta \alpha_k^{\text{optim}} (t) = & ~ \frac{ \sqrt{7}}{6}[f_{e}^{k}(t) e_k(t) + f_{e}^{i}(t) e_i (t) + f_{e}^{j} (t) e_j(t)]    + \frac{1}{2 \sqrt{3} a_{\text{o}} }[\delta a^{s}_{ik}(t) + \delta a^{s}_{jk}(t)] +  \frac{1}{2} \delta \lambda ^{s}_{ji}(t) +\frac{1}{2} \cos \overline{i}_{\text{o}} \, \delta \Omega^{s}_{ji}(t), \label{eq:delta_alphaij_optim}
\end{align}
\end{widetext}
with right-hand functions adopting the conditions given by Eq.~(\ref{eq:optim_con}), i.e., $  a_{\text{o}} = \overline{a}_{k}^{\, \text{optim}} $, $ \overline{i}_{\text{o}}  = \overline{i}_{k}^{\, \text{optim}}  $, $ e_k(t) = e^{\text{optim}}_k(t) $, etc. Equations (\ref{eq:delta_Lij_optim})-(\ref{eq:delta_alphaij_optim}) reveal the optimized TianQin triangle as intrinsic fluctuation variations induced by perturbations, with amplitude dependent on eccentricities and short-period variations in other elements. Notably, conditions in Eq.~(\ref{eq:optim_con}) correspond to the optimal stable configuration, and therefore can provide useful guidelines for numerical optimization and orbit control.


\section{\label{sec:4} Concluding Remarks}

Detecting GWs with TianQin requires a stable three-satellite constellation, configured as closely to an equilateral triangle as possible. In high Earth orbits, gravitational perturbations, especially from lunar and solar influences, can distort this configuration. To quantify the impact, we have developed an analytical model delineating the effects of lunar and solar point masses on the TianQin constellation. This model provides expressions for three kinematic indicators, including arm-lengths, relative velocities, and breathing angles, derived from the first-order perturbation solution for satellite orbital elements.

The analysis of these indicators has revealed that gravitational perturbations induce secular, long-period, and short-period variations in satellite orbital elements, leading to relative motion between satellites and distortions in the constellation. These distortions appear as three distinct types, i.e., linear drift, bias, and fluctuation. The drift, progressively increasing over time, is a primary destabilizing factor affecting arm-lengths and breathing angles but has almost no impact on relative velocities. To alleviate design constraints on onboard scientific payloads, these three distortions have been further optimized. It is demonstrated that both drift and bias can be eliminated, and fluctuation amplitude reduced, if the three orbits adhere to the following constraints on average: synchronized orbital periods, aligned orbital planes, equally spaced phases, and minimized eccentricities. The expressions for the optimized indicators are presented, revealing that the optimized TianQin constellation displays only fluctuation with amplitude dependent on eccentricities and short-period variations in other elements.

These results can provide valuable insights and guidelines for enhancing the stability of the GW observatory constellation, such as in numerical optimization and orbit control. For future works, this model will be extended to incorporate the influence of initial orbit errors \cite{Qiao2023a,Zhou2022,Jia2022,An2022}. The perturbation solution can be applied to explore the dynamics of TianQin satellite eccentricity, closely linked to the constellation stability. Potential applications in celestial mechanics, especially for high-inclination TianQin-like orbits subject to the Kozai–Lidov effect \cite{Kozai1962,Lidov1962}, may arise. Further discussions are deferred to future work.


\begin{acknowledgments}
The authors thank Jianwei Mei, Yunhe Meng, Defeng Gu, Liang-Cheng Tu, Cheng-Gang Shao, Yan Wang, and Jun Luo for helpful discussions and comments. Special thanks to the anonymous referee for valuable suggestions and comments. X. Z. is supported by the National Key R\&D Program of China (Grant Nos.~2020YFC2201202 and 2022YFC2204600), NSFC (Grant No.~12373116), and Fundamental Research Funds for the Central Universities, Sun Yat-sen University (Grant No. 23xkjc001). 
\end{acknowledgments}


\appendix

%

\section{Table of Symbols}\label{Appendix:symbols}

Table \ref{table:symbols} below lists the main symbols used in the paper and their meanings for quick look-ups.

\setlength\LTleft{0pt}
\setlength\LTright{0pt}
\begin{longtable}{@{\extracolsep{0in}}p{1.1in}p{2.2in}}
\caption{\label{table:symbols} List of symbols and their meanings.}\\*
\hline
\hline
\noalign{\vspace{3pt}}%
Symbols & \multicolumn{1}{c}{Meanings}\\*[3pt]
\hline 
\endfirsthead

\multicolumn{2}{l}{TABLE~\ref{table:symbols}. (\textit{Continued})}
\rule{0pt}{12pt}\\[3pt]
\hline
\hline
\noalign{\vspace{3pt}}%
Symbols & \multicolumn{1}{c}{Meanings}\\*[3pt]
\hline 
\noalign{\vspace{3pt}}%
\endhead
\noalign{\nobreak\vspace{3pt}}%
\colrule
\multicolumn{2}{r}{\textit{(Table continued)}}
\endfoot
\noalign{\nobreak\vspace{3pt}}%
\botrule
\endlastfoot
$ t $            & Time \\
$ t_{0} $        & Reference epoch \\
UTC              & Coordinated Universal Time \\
$ a $            & Semimajor axis \\
$ e $            & Orbital eccentricity \\
$ i $            & Orbital inclination \\
$ \Omega $       & Longitude of ascending node \\
$ \omega $       & Argument of perigee \\
$ \nu $          & True anomaly \\
$ M $            & Mean anomaly \\
$ E $            & Eccentric anomaly \\
$ \text{u} \,( = \omega + \nu ) $ & Argument of latitude \\
$ \lambda $    & Defined as $ (\omega + M)  $ \\
$ \xi \,( = e\cos \omega ) $      & Singularity-free variable for eccentricity \\
$ \eta \,( = e\sin \omega ) $     & Singularity-free variable for eccentricity \\
$ \sigma $       & General representation of orbital elements \\
$ \sigma_{0} $   & Initial orbital elements \\
Subscript e      & Refers to the Earth \\
Subscript s      & Refers to the Sun \\
Subscript m      & Refers to the Moon \\
Subscript $ J_{2} $ & Refers to the Earth's oblateness $ J_{2} $ \\
Subscript 0      & Denoting initial values or zeroth order \\
Subscript 1      & Denoting SC1 or first order \\
Subscript 2      & Denoting the Sun, SC2 or second order \\
Subscript 3      & Denoting the Moon or SC3 \\
Subscript $\text{o}$ & Denoting nominal values \\
Notation $ c $ & Denoting secular variation \\ 
Notation $ l $ & Denoting long-period variation \\
Notation $ l[c] $ & Denoting special long-period variation \\
Notation $ l[l] $ & Denoting general long-period variation \\
Notation $ s $ & Denoting short-period variation \\
$ i $, $ j $, $ k $ &  Represent satellites and take values of 1, 2, 3  \\ 
$ \kappa $, $ p $, $ q $ & Represent components of the perturbation solution \\
$ \dot{y} $      & Time derivative of $ y $ \\
$\overline{y}$   & Mean orbital elements or average values \\
$ \Delta y $     & Difference (See Eq.~(\ref{eq:nomOrb_real})) \\
$ \delta y $     & Change relative to the nominal value $ y_{\text{o}} $ (See Eqs.~(\ref{eq:Lij_decompose}), (\ref{eq:orbit_same_k-vs-nomOrb_real}), and (\ref{eq:delta_orb})) \\ 
$ \delta_{\!\Delta} y  $  &  See Eq.~(\ref{eq:delta_Delta}) \\
$ \mu \,(= G M_{\text{e}}) $ & Gravitational constant of the Earth \\
$ \mu _2 \,(= G M_{\text{s}}) $ & Gravitational constant of the Sun \\
$ \mu _3 \,(= G M_{\text{m}}) $ & Gravitational constant of the Moon \\
$ R_{\text{e}} $ & Equatorial radius of the Earth   \\
$ J_2 $           & Second zonal harmonic coefficient \\
$ \epsilon $      & Obliquity of the ecliptic \\
$ U $             & Gravitational potential \\
$ U_{0} $         & Central gravitational potential \\
$ \mathcal{R} $   & Perturbative potential \\ 
$ P_{N}(x) $      & Legendre polynomial of degree $N$ \\
$ \mathcal{N} $   & Truncation degree \\
$ R_{x}(\gamma ) $, $ R_{z}(\gamma ) $ & Rotation matrices about the $ x $ and $ z $ axes by angle $ \gamma $ \\
$ f_{0} $, $ f_{1} $ & Zeroth and first order of the function $ f $ \\ 
$ \delta _{\iota} $ & Takes values 0 or 1 (see Eqs.~(\ref{eq:delta_01_a}) and (\ref{eq:delta_01})) \\
$ r $             & Geocentric satellite distance \\
$ \mathbf{r} $    & Geocentric satellite position vector \\
$ \mathbf{\dot{r}} $    & Geocentric satellite velocity vector \\
$ \mathbf{\hat{r}} $, $ \mathbf{\hat{r}_{2}} $, $ \mathbf{\hat{r}_{3}} $ & Unit vectors for satellite, the Sun, and the Moon, respectively \\
$ \psi_{2}, \psi_{3} $      & Geocentric angles between satellite and the Sun, and the Moon \\
$ \varphi $       & Geocentric latitude in Earth-fixed coordinate system \\
$ r_2 $           & Sun-Earth distance \\
$ \overline{r}_2 $ & Mean Sun-Earth distance \\
$ \text{u}_{2}$   & Ecliptic longitude of the Sun \\
$ u_{2}$          & Mean longitude of the Sun \\
$  U_{2}  $       & Defined as $ (u_{2} - \overline{\Omega}) $ \\
$ r_3 $           & Earth-Moon distance \\
$ \overline{r}_{3} $ & Mean Earth-Moon distance \\
$ \overline{i}_{3} $ & Mean inclination of the Moon's orbit \\
$  \overline{\Omega}_{3}$ & Secular variation in the Moon's longitude of ascending node \\
$ \overline{M}_{3}  $ & Secular variation in the Moon's mean anomaly \\
$ \text{u}_{3}  $ & Latitude argument of the Moon \\
$ u_{3} $         & Secular variation in the Moon's latitude argument \\
$ \Delta u_{3} $  & Defined as $ (u_{3} - \overline{M}_{3}) $ \\
$  \theta _{3} $  & Defined as $ (\overline{\Omega} - \overline{\Omega} _{3}) $ \\
$ n $             & Mean motion of satellite  with an orbit period of 3.64 days \\
$ n_{\Omega}' $  & Rate of change of $ \Omega $ \\
$ n_{ \lambda }' $ & Rate of change of $ \lambda $ \\
$ n_{2} $         & Mean motion of the Earth with a period of 365.2564 days \\
$n_{3}$           & Rate of change of $ u_{3}$ with a period of 27.21 days (draconic month) \\ 
$n_{\! M_3}$      & Rate of change of $ \overline{M}_{3}$ with a period of 27.55 days (anomalistic month) \\
$ n_{\Omega_{3}} $ & Rate of change of $ \overline{\Omega}_{3}$ with a period of 18.6 years \\
$ \Delta n_{3} \,(= n_{3} - n_{\!M_3}) $ & Rate of change of $ \Delta u_{3}$ with a period of 6.0 years \\
$n_{\theta _3}\,(= n_{\Omega}'-n_{\Omega _3})$ & Rate of change of $ \theta _{3}$ \\
$n_{U_{2}}\,(=n_{2} - n_{\Omega}') $ & Rate of change of $ U_{2}$ \\
$ L_{ij} $        & Arm-length formed by SC$ i $ and SC$ j $ \\
$ v_{ij} $        & Relative line-of-sight velocity (rate of change of $ L_{ij} $) \\
$ \alpha_{k} $    & Breathing angle at SC$ k $ \\
$ L_\text{o} $, $ v_\text{o} $, $ \alpha_\text{o} $ & Nominal values of arm-length, relative velocity, and breathing angle \\
$ f^{\sigma(N)}_{[q]}(i), f^{\sigma(N)}_{[p,\, q]}(i)  $ & Inclination functions in the lunar perturbation solution \\ 
\end{longtable}

\section{Model derivation and verification}
\label{sec:Ana_solution_DerivVerif12} 

\subsection{Derivation and motivation of Eq.~(\ref{eq:R_m})}
\label{sec:Ana_solution_DerivVerif0} 

For lunar point-mass perturbation, the perturbative potential $ \mathcal{R}_{\text{m}} $ is represented as \cite{Liu2000}
\begin{align}
\mathcal{R}_{\text{m}} &=  \mu _{3} \left ( \frac{1}{|\mathbf{r} - \mathbf{r_{3}}|} - \frac{\mathbf{r_{3}}}{r_{3}^{3}}\cdot \mathbf{r}   \right ) \\
&= \mu _{3}  \left ( \frac{1}{ \sqrt{{r}^{2}- 2 \, r \, r_{3}  \cos \psi _{3} + r_{3}^{2}}} - \frac{r}{r_{3}^{2}} \cos \psi _{3} \right ) \label{eq:R_m_2},
\end{align}
where $ r=\frac{a(1-e^{2})}{1 + e \cos \nu } $, $ r_{3}=\frac{a_{3}(1-e_{3}^{2})}{1 + e_{3} \cos \nu_{3} } $, and $ \cos \psi_{3} = \mathbf{\hat{r}_{3}} \cdot   \mathbf{\hat{r}} $. $ \mathcal{R}_{\text{m}} $, described by Eq.~(\ref{eq:R_m_2}), exhibits a square root form, introducing challenges in solving the Lagrange perturbation equations. This complexity can be circumvented by expressing $ \frac{1}{|\mathbf{r} - \mathbf{r}_{3}|} $ as an expansion of Legendre polynomials:
\begin{align}
 \frac{1}{|\mathbf{r} - \mathbf{r_{3}}|} & = \frac{1}{r_{3}} \left [ 1- 2\left ( \frac{r}{r_{3}}\right ) \cos \psi _{3} + \left ( \frac{r}{r_{3}} \right )^{2}    \right ]^{-1/2} \label{eq:R_m_3}\\
& =\frac{1}{r_{3}}\sum_{N=0}^{ \infty  }\left (\frac{r}{r_{3}} \right )^{N} P_{N}(\cos \psi_{3} ).  \label{eq:R_m_4}
\end{align}
Further substituting Eq.~(\ref{eq:R_m_4}) into Eq.~(\ref{eq:R_m_2}) and removing the first term $ \frac{1}{r_{3}} $ (as $\frac{\partial}{\partial \sigma} ( \frac{1}{r_{3}} ) = 0$ after substitution into Lagrange's equations) yields 
\begin{align}
\mathcal{R}_{\text{m}} = \frac{ \mu _{3} }{r_{3}}\sum_{N=2}^{ \infty }\left (\frac{r}{r_{3}} \right )^{N} P_{N}(\cos \psi_{3} ).\label{eq:R_m6} 
\end{align}
This formulation proves more advantageous for solving the Lagrange equations than Eq.~(\ref{eq:R_m_2}). Based on estimated magnitudes and validation through numerical simulations, we set the maximum degree of Legendre polynomials in $ \mathcal{R}_{\text{m}} $ at $ N=6 $. Additionally, the solar potential $ \mathcal{R}_{\text{s}} $ resembles Eq.~(\ref{eq:R_m6}), with the maximum degree set at $ N=2 $.

\subsection{Derivation of Eqs.~(\ref{eq:sol_form1})-(\ref{eq:sol_perturbed})}
\label{sec:Ana_solution_DerivVerif1} 

The analytical expressions for Eqs.~(\ref{eq:sol_form1})–(\ref{eq:sol_perturbed}) can be derived by applying perturbation methods to solve Eq.~(\ref{eq:perturbation}). To enhance the accuracy of the analytical solution, it is more advantageous to use the mean orbital elements $ \overline{\sigma} (t) $ \cite{Kozai1973,Liu2000}, corresponding to a long-term precessing elliptical orbit, rather than the Keplerian orbit $ \sigma^{(0)} (t) $ as the reference solution. Consequently, the perturbation solution's form (\ref{eq:sol_form1}) is reformulated as
\begin{align}
\sigma (t) = \overline{\sigma} (t) + [ \sigma _{l[l]}(t) +\sigma _{s}(t) ]\label{eq:sigmat_split}
\end{align}
with
\begin{align}
\overline{\sigma} (t)  = \overline{\sigma}^{(0)} (t)  + \Delta \sigma _{c}(t) + \Delta \sigma _{l[c]}(t), \label{eq:sigmaPJt_NiPJ}
\end{align}
and
\begin{align} 
& \overline{\sigma}^{(0)} (t)=\overline{\sigma}_{0} +\delta_{\iota} \overline{n}(t-t_{0}), \\
& \overline{\sigma}_{0} = \sigma_{0} - [\sigma _{l[l]}(t_{0})+  \sigma _{s}(t_{0}) ], \label{eq:sigmaPJt0_NiPJ}
\end{align}
where $ \overline{\sigma}^{(0)} $ represents the unperturbed secular variations, $ \overline{\sigma}_{0} $ is the initial mean elements, $ \Delta \sigma_{c} (t):= \sigma_{c} (t) - \sigma_{c} (t_{0}) $, and $ \Delta \sigma_{l[c]} (t):= \sigma_{l[c]} (t) - \sigma_{l[c]} (t_{0}) $. Notably, $ \sigma_{l[c]} (t) $ is incorporated into $ \overline{\sigma} (t) $, considering its short-term behavior akin to secular variation.

In relation to the left-side partitioning of Eq.~(\ref{eq:perturbation}) concerning $ \sigma (t) $, the function $ f_{1} $ on the right is similarly decomposed into
\begin{align}
 f_{1} = f_{1c} +f_{1l[c]} +f_{1l[l]} +f_{1s}.\label{eq:f1_split}
\end{align}
$ f_{1c} $ depends solely on $ \overline{a} $, $ \overline{\xi} $, $ \overline{\eta} $, and $ \overline{i} $. Both $f_{1l[c]}$ and $f_{1l[l]}$ involve trigonometric functions with arguments related to slow variables, such as $ \Omega_{3} $ with an 18.6-year period and $ u_{3} $ with a 27.21-day period, while $f_{1s}$ incorporates the fast variable $ \lambda $, which has a 3.64-day period, as the argument. The decomposition in Eq.~(\ref{eq:f1_split}) is achieved through averaging \cite{Smith1962,Prado2003,Roscoe2013}, where, for instance, $ f_{1s} $ is obtained via
\begin{align}
f_{1s}= f_{1} - \left \langle f_{1} \right \rangle_{\lambda},\qquad \left \langle f_{1} \right \rangle_{\lambda}  := \frac{1}{2\pi} \int_{0}^{2\pi }f_{1} \, d \lambda. \label{eq:PJ_x1}		
\end{align}				
A similar averaging over slow variables is applied to derive $ f_{1c} $, $ f_{l[c]} $, and $ f_{l[l]} $. Moreover, given that the TianQin orbits are nearly circular with $ \overline{e} \simeq 0.0005 $ \cite{Ye2019}, the terms on the right side of Eq.~(\ref{eq:f1_split}) consider only the leading-order effects of eccentricity for simplicity.

By inserting the formal solution (\ref{eq:sigmat_split}) into both sides of Eq.~(\ref{eq:perturbation}) and conducting a Taylor expansion around $ \overline{\sigma} (t) $, the comparison of coefficients for the same powers ($ \varepsilon^{0} $, $ \varepsilon^{1} $, $ \varepsilon^{2} $, $ \cdots $) yields \cite{Liu2000}
\begin{align}
& \overline{\sigma}^{(0)}(t) = \int^{t}_{t_{0}} [f_{0}]_{\overline{a}}\,dt = \overline{\sigma}_{0} + \delta_{\iota} \overline{n}(t-t_{0}),\label{eq:sigma_c0} \\
&  \sigma _{c}^{(1)}(t) = \int^{t} [f_{1c}]_{\overline{\sigma}}\,dt ,\label{eq:sigma_c1}\\
& \sigma _{l[c]}^{(1)}(t) = \int^{t} [f_{1l[c]}]_{\overline{\sigma}}\,dt ,\label{eq:sigma_lc1} \\
&  \sigma _{l[l]}^{(1)}(t) = \int^{t} [f_{1l[l]}]_{\overline{\sigma}}\,dt ,\label{eq:sigma_l1}\\
&  \sigma _{s}^{(1)}(t) = \int^{t} [\frac{\partial  f_{0}}{\partial a}a _{s}^{(1)}+ f_{1s}]_{\overline{\sigma}}\,dt ,\label{eq:sigma_s1} \\
&  \sigma _{c}^{(2)}(t) = \int^{t} \biggl[\frac{1}{2}\frac{\partial^2 \! f_{0}}{\partial a^2}[a _{s}^{(1)}]^{2}_{c} \nonumber \\
& \qquad \quad \;\;\,\, +\biggl( \sum_{j=1}^{6} \frac{\partial  f_{1}}{\partial \sigma_{j}}[\sigma _{s}^{(1)}+\sigma _{l[l]}^{(1)}]_{j}\biggr)_{\!\!c}  \biggr]_{\! \overline{\sigma}}\,dt, \label{eq:order2} \\
& \cdots.  \nonumber 
\end{align}
The superscript in parentheses denotes the order of the perturbation solution; in this paper, we focus on the first-order solution. Utilizing Eqs.~(\ref{eq:sigma_c1})-(\ref{eq:sigma_s1}), we derive explicit expressions for the four components of $ \sigma^{(1)} (t) $, presented in Appendix \ref{sec:Ana_solution_expression}. Particularly, for $ \xi_{c} $, $ \xi_{l[c]} $, $ \eta_{c} $, and $ \eta_{l[c]} $, it is more reasonable to directly solve the oscillation equations they satisfy (see Eqs.~(\ref{eq:dxidt_ms-c}) and (\ref{eq:detadt_ms-c})) \cite{Liu2000}; detailed derivations are provided in Appendix \ref{sec:Ana_solution_lunisolar}.

Note that in Eqs.~(\ref{eq:sigma_c0})-(\ref{eq:order2}), $ \sigma $ on the right-hand side all take the form $ \overline{\sigma}(t) $ defined in Eq.~(\ref{eq:sigmaPJt_NiPJ}). For $ i $, $ \Omega $, and $ \lambda $, they are embedded in the trigonometric functions of $ f_{1} $. To enable integrable solutions, $ \sigma_{l[c]}(t) $ in $ \overline{\sigma}(t) $ is approximated as a linear term with a rate of change $ \widetilde{n}_{\sigma} $, 
\begin{align}
\widetilde{n}_{\sigma}  = \frac{ \left \langle \Delta \sigma _{l[c]}(t)  \right \rangle_{\tau} }{\tau},\qquad \left \langle f(t) \right \rangle_{\tau} := \frac{1}{\tau}\int_{t_{0}}^{t_{0}+\tau} f(t) \, dt, 
\end{align}
where $ \tau $ denotes the duration, implying
\begin{equation}
\overline{\sigma}(t) \simeq  \left\{
\begin{array}{ll}
 \overline{\sigma }_{0} + n_{\sigma}'(t -t_{0}),   & \sigma \in \{\Omega ,\lambda \}, \smallskip\\
  \left \langle  i (t) \right \rangle_{\tau},   & \sigma = i, \\
\end{array} 
\right.\label{eq:sigma_niPJ_msJ2_simplify}
\end{equation}
with
\begin{align}
n_{\sigma}' :=  \delta_{\iota}  \overline{n} + n_{\sigma}+ \widetilde{n}_{\sigma}.\label{eq:n_sigma_niPJ} 
\end{align}
Here, $ n_{\sigma}  $ represents the rate of change for $ \sigma_{c}(t) $, given by $ n_{\sigma} = n_{\sigma \text{s}}+n_{\sigma \text{m}}+n_{\sigma \! _{J_{2}}} $. Additionally, $ \overline{i}(t) $ is approximated as the mean value $ \left \langle i (t) \right \rangle_{\tau} $, due to its small secular variation.

\subsection{Verification of Eqs.~(\ref{eq:sol_form1})-(\ref{eq:sol_perturbed})}
\label{sec:Ana_solution_DerivVerif2} 

To validate the derived analytical solution for satellite orbits, we conduct high-precision numerical orbit simulations using the NASA General Mission Analysis Tool (GMAT) \cite{GMAT}. The force models, consistent with those in Ref.~\cite{Ye2019}, include the point-mass gravity fields of the Moon, Sun, and solar system planets (the ephemeris DE421), a $10\times10$ spherical-harmonic model of the Earth's gravity field (JGM-3), and the first-order relativistic correction. Non-gravitational perturbations, such as solar radiation pressure, are omitted as the satellites are drag-free controlled. Additionally, an adaptive step, ninth-order Runge-Kutta integrator with eighth-order error control (RungeKutta89) is employed, with the maximum integration step size set to 45 minutes. Initial orbital elements for the test orbits are detailed in Table \ref{table:Ini_Elements_TestAna}. Orbit-1 corresponds to the nominal orbit of the TianQin satellite. In addition, three cases with different inclinations are
considered to facilitate a more comprehensive validation, considering that the inclination is a crucial parameter in the analytical solution.

\begin{table}[ht]
  \caption{Initial orbital elements of the test orbits in the J2000-based Earth-centered ecliptic coordinate system at the epoch 1 Jan, 2034, 00:00:00 UTC.}
\begin{ruledtabular}
\begin{tabular}{ccccccc}
Test orbits & $ a_{0} $\,(km) &  $ e_{0} $ & $ i_{0} $\,($ ^{\circ} $)  & $ \Omega_{0} $\,($ ^{\circ} $) & $ \omega_{0} $\,($ ^{\circ} $) & $ \nu_{0} $\,($ ^{\circ} $) \\ 
\hline
Orbit-1 & 100\,000 & 0 & 94.7  & 210.4 & 0 & 60 \\
Orbit-2 & 100\,000 & 0 & 65.0  & 210.4 & 0 & 60 \\
Orbit-3 & 100\,000 & 0 & 35.0  & 210.4 & 0 & 60 \\
Orbit-4 & 100\,000 & 0 & \hphantom{0}5.0  & 210.4 & 0 & 60 \\
\end{tabular}
\end{ruledtabular}
    \label{table:Ini_Elements_TestAna}
\end{table}

The comparison between analytical and numerical orbits reveals the errors $ \Delta \sigma(t) := \sigma_{\text{ana}}(t) - \sigma_{\text{num}} (t) $ ($ \sigma \in \{a,e,i, \Omega,\omega, \lambda \} $). Statistical results, shown in Table \ref{table:error_statistics}, demonstrate that analytical expressions for $ e $, $ i $, $ \Omega $, and $ \lambda $ are in good agreement with numerical simulations, with the relative mean deviation of $ e $ being less than $6\%$, and long-term deviations for $ i $, $ \Omega $, and $ \lambda $ being small. In addition, there are relatively large errors in $ a $ and $ \omega $, with the latter having a minor influence on the constellation stability (see, e.g., Eq.~(\ref{eq:delta_Lij})). Table \ref{table:error_statistics} also includes a comparison of satellite positions, denoted as $ | \Delta \mathbf{r}(t) | :=  | \mathbf{r}_{\text{ana}}(t) - \mathbf{r}_{\text{num}} (t) | $. For the TianQin orbit, the average and maximum deviations over a 5-year period are approximately 87 km and 210 km, respectively.

\begin{table*}[htp]
  \caption{Comparison of the analytical solution for satellite orbits with numerical simulations over 5 years, indicating mean errors (and maximum errors). To address the secular variation of $ \Omega $, the second-order solution $ \Omega_{c}^{(2)}(t) $ obtained from Eq.~(\ref{eq:order2}) has been incorporated into the analytical solution. When propagating $ \lambda $ and, consequently, $ \mathbf{r} $ using Eqs.~(\ref{eq:sol_form1})-(\ref{eq:sol_perturbed}), the mean semimajor axes of the numerical orbits were employed.}
\begin{ruledtabular}
\begin{tabular}{cccccccccc}
  Test orbits   &  $ \Delta a $\,(km)  &  $  \Delta e /e $\,(\%)&  $ \Delta i $\,($ ^{\circ} $)    & $ \Delta \Omega $\,($ ^{\circ} $) & $ \Delta \omega $\,($ ^{\circ} $)  &   $ \Delta \lambda  $\,($ ^{\circ} $)  &  $ | \Delta \mathbf{r} |  $\,(km) \\
        \hline
 Orbit-1 &  $-$1.4 ($-$16.1) & $+$5.7 ($+$7.3) & $-$0.00 ($-$0.02) &  $-$0.00 ($-$0.01) & $-$1.1 ($+$29.9) & $-$0.05 ($-$0.11)  & \hphantom{0}87 (210) \\
 Orbit-2 &  $+$0.4 ($-$15.5) & $+$1.2 ($+$3.0) & $+$0.01 ($+$0.06) &  $+$0.08 ($+$0.12) & $-$1.1 ($+$70.9) & $+$0.01 ($-$0.10)  & 126 (240) \\
 Orbit-3 &  $+$1.9 ($+$16.2) & $-$4.4 ($-$8.5) & $+$0.04 ($+$0.14) & $+$0.08 ($+$0.14) & $-$1.6 ($-$44.4) & $+$0.08 ($+$0.18) &  291 (502) \\
 Orbit-4 &  $+$2.2 ($+$18.2) & $-$1.0 ($+$4.9) & $-$0.02 ($-$0.05) & $-$0.38 ($-$0.96) & $+$7.3 ($+$19.9) & $+$0.42 ($+$1.02)  & 161 (426) \\  
    \end{tabular}
    \end{ruledtabular}
    \label{table:error_statistics}
\end{table*}

For future improvements, potential dominant sources causing the aforementioned errors are briefly outlined as follows. Firstly, simplified analytical coordinates for the Sun and Moon (see Sec.~\ref{sec:MS_parameter_simplify}) were utilized, instead of higher-precision ones with multiple trigonometric corrections \cite{Liu2000,Montenbruck2001}. Secondly, smaller perturbative effects, including those from other planets in the solar system and the nonspherical gravitational field of the Sun and Moon, were omitted in Eq.~(\ref{eq:R_smJ2}). Lastly, the second-order solution was lacking, and the next-leading-order eccentricity effect was neglected in Eq.~(\ref{eq:f1_split}), etc.

\subsection{Model verification for the three indicators}
\label{sec:Ana_solution_DerivVerif3} 

Appendix \ref{sec:Ana_solution_DerivVerif2} verifies the analytical solution for satellite orbits, focusing on individual satellites. Additionally, this subsection presents the verification of the analytical expressions for the kinematic indicators of the three-satellite constellation: $L_{ij}(t)$, $v_{ij}(t)$, and $\alpha_{k}(t)$, derived from Eq.~(\ref{eq:sigma_msJ2}) (or Eqs.~(\ref{eq:sol_form1})-(\ref{eq:sol_perturbed})).

\begin{figure*}[htb]
\begin{minipage}{0.46\textwidth} 
\flushright 
\includegraphics[width=\textwidth]{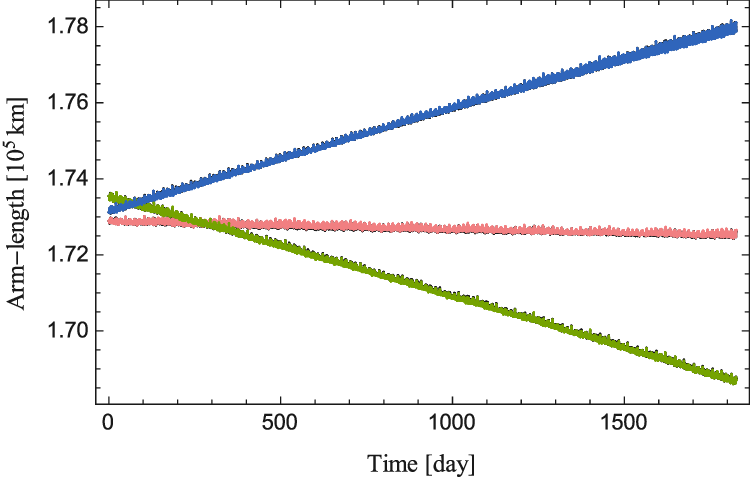}
\end{minipage} 
\hspace{0.2cm}
\begin{minipage}{0.46\textwidth} 
\flushright 
\includegraphics[width=0.992\textwidth,height=0.655\textwidth]{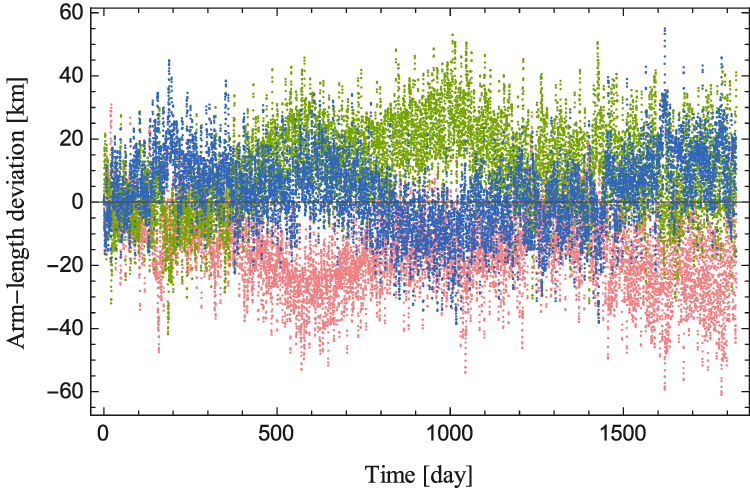}
\end{minipage} 
\begin{minipage}{0.46\textwidth} 
\flushright 
\includegraphics[width=0.975\textwidth,height=0.645\textwidth]{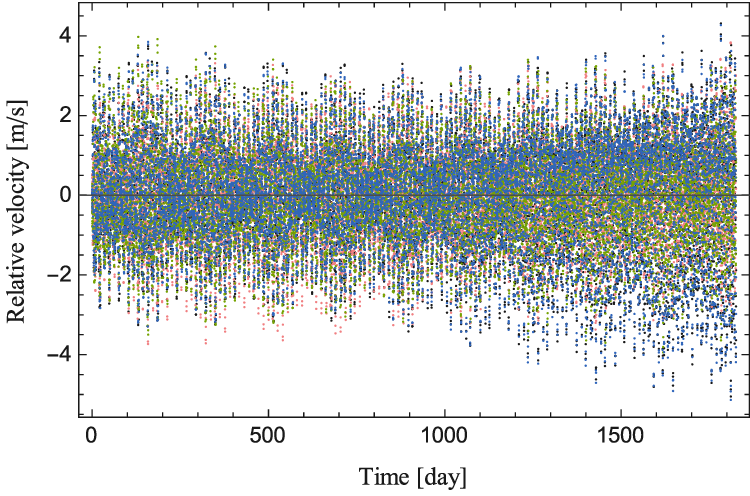}
\end{minipage}
\hspace{0.2cm}
\begin{minipage}{0.46\textwidth} 
\flushright 
\includegraphics[width=\textwidth]{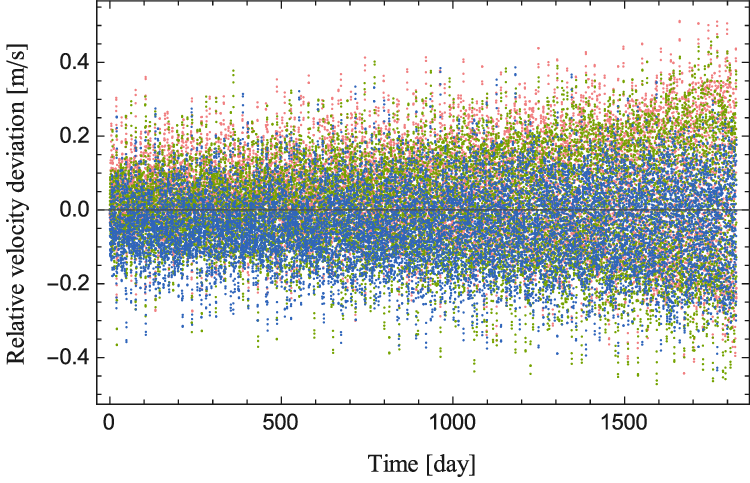}
\end{minipage} 
\begin{minipage}{0.46\textwidth} 
\flushright 
\includegraphics[width=0.975\textwidth,height=0.645\textwidth]{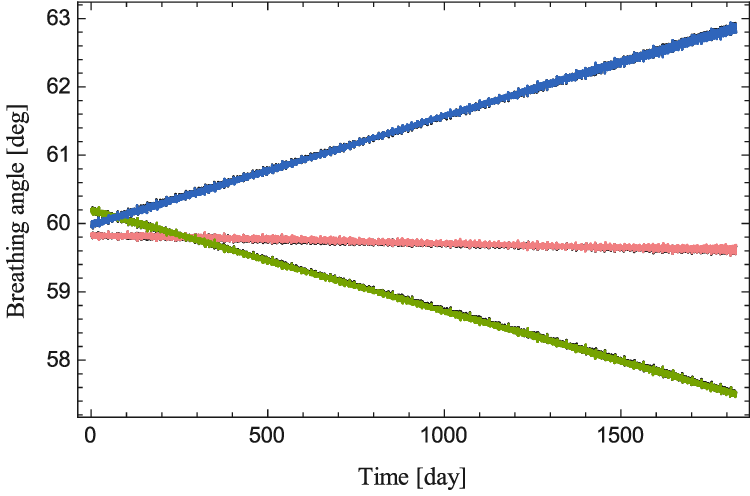}
\end{minipage}
\hspace{0.2cm}
\begin{minipage}{0.46\textwidth} 
\flushright 
\includegraphics[width=\textwidth,height=0.665\textwidth]{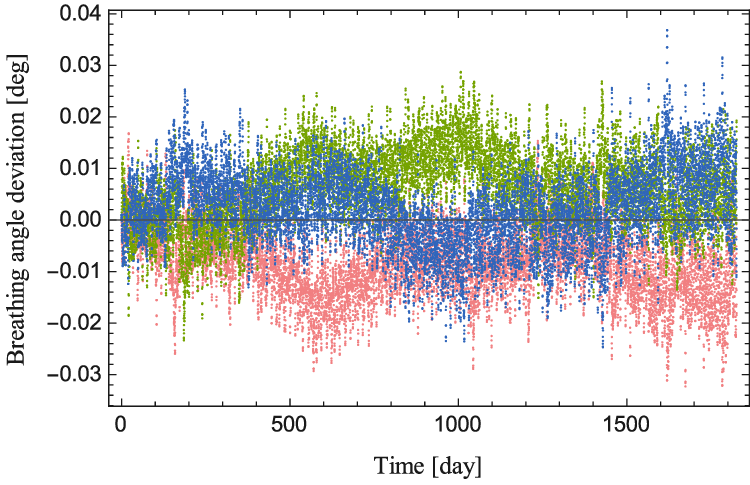}
\end{minipage} 
\caption{\label{fig:evolu_anaVSnum} 
Time evolution of three kinematic indicators for the analytical model compared to the numerical model. The left panel displays subplots illustrating variations in arm-lengths ($ L_{ij} $), relative velocities ($ v_{ij} $), and breathing angles ($ \alpha_{k} $), respectively. Colors in each subplot represent the numerical model (blue for $ L_{12} $, $ v_{12} $, or $ \alpha_{3} $; red for $ L_{23} $, $ v_{23} $, or $ \alpha_{1} $; green for $ L_{31} $, $ v_{31} $, or $ \alpha_{2} $), while black denotes the analytical model (only $ v_{12} $ is plotted for clarity in relative velocities). The right panel illustrates the temporal evolution of discrepancies between the analytical and numerical models for these three indicators.}
\end{figure*}

The time evolution of these three indicators in both analytical and numerical models is plotted in the left panel of Fig.~\ref{fig:evolu_anaVSnum} for a representative set of initial orbital elements provided in Table \ref{tab:orbits_error}. In the numerical model, the considered perturbations, integrator, and step size align with those detailed in Appendix \ref{sec:Ana_solution_DerivVerif2}. The right panel illustrates the time evolution of the deviations between the analytical and numerical models for these three quantities. Figure \ref{fig:evolu_anaVSnum} indicates that the analytical model can effectively capture the long-term variations in the indicators, while noticeable periodic deviations exist. Numerical simulation results suggest that these deviations primarily arise from approximations in the Sun and Moon analytical dynamical model. By employing higher-precision models for solar and lunar motion \cite{Liu2000,Montenbruck2001}, incorporating numerous trigonometric correction terms in the Sun's ecliptic longitude $ \text{u}_{2} $ and the Moon's latitude argument $ \text{u}_{3} $, these deviations can be effectively reduced. On the other hand, this enhanced complexity presents challenges in analytically solving the Lagrange equations, as $ \text{u}_{2} $ and $ \text{u}_{3} $ themselves involve trigonometric functions (see Eqs.~(\ref{eq:r_unit_vec}) and (\ref{eq:r3_unit_vec})). Additionally, beyond the orbits specified in Table \ref{tab:orbits_error}, the analytical model has been validated on two additional sets: nominal orbits (with SC1's initial elements matching those of Orbit-1 in Table \ref{table:Ini_Elements_TestAna}) and optimized orbits (refer to Table 3 in \cite{Ye2019}), yielding consistent results.

\begin{table}[ht]
\caption{\label{tab:orbits_error} Initial orbital elements for simulated TianQin orbits in the J2000-based Earth-centered ecliptic coordinate system at the epoch of 22 May, 2034, 12:00:00 UTC. These initial elements deviate by approximately 1 km, $ 10^{-5} $, 0.2$ ^{\circ} $, 0.2$ ^{\circ} $, 0.1$ ^{\circ} $, and 0.1$ ^{\circ} $ in $a_{0}$, $e_{0}$, $i_{0}$, $\Omega_{0}$, $\omega_{0}$, and $\nu_{0}$, respectively, from the optimized orbits (cf.~Table 3 in \cite{Ye2019}). The subsequent orbital evolution is illustrated in Fig.~\ref{fig:evolu_anaVSnum}.}
\begin{ruledtabular}
\begin{tabular}{ccccccc}
 & $ a_{0} $\,(km) &  $ e_{0} $ & $ i_{0} $\,($ ^{\circ} $) \\ 
\hline
SC1 & \hphantom{0}99\,996.572\,323 & 0.000\,440 & 94.897\,997  \\
SC2 & 100\,010.400\,095 & 0.000\,010 & 94.904\,363 \\
SC3 & \hphantom{0}99\,992.041\,899 & 0.000\,296 &  94.509\,747 \\
 & $ \Omega_{0} $\,($ ^{\circ} $) & $ \omega_{0} $\,($ ^{\circ} $) & $ \nu_{0} $\,($ ^{\circ} $) \\
SC1 & 210.645\,892 & 358.724\,463 & \hphantom{0}61.429\,603 \\
SC2 & 210.240\,199 & 359.900\,000 & 180.130\,706 \\
SC3 & 210.644\,582 & 359.901\,624 & 299.812\,164 \\
\end{tabular}
\end{ruledtabular}
\end{table}

Furthermore, the expressions (\ref{eq:delta_Lij}), (\ref{eq:delta_vij}), and (\ref{eq:delta_alphaij}), derived from the series expansion of the three indicators, have been verified. The results suggest that to achieve a deviation magnitude similar to that before the series expansion, the second-order term $ \delta L_{ij}(\delta \lambda(t)^{2}) = \frac{\sqrt{3}}{4} a_{\text{o}} \delta \lambda_{i}(t) \delta \lambda_{j}(t) - \frac{\sqrt{3}}{8} a_{\text{o}} [\delta \lambda_{i}(t)^{2} + \delta \lambda_{j}(t)^{2}] $ in the arm-length, where $ \delta \lambda(t)^{2} \sim \delta n^{2} \, t^{2} $ rapidly increases with time, needs to be taken into account. Notably, for the relative velocity and breathing angle, these second-order terms are both zero, $ \delta v_{ij}(\delta \lambda(t)^{2}) = \delta \alpha_{k}(\delta \lambda(t)^{2}) = 0 $.

\section{Explicit expressions for terms in Eq.~(\ref{eq:sol_perturbed})}
\label{sec:Ana_solution_expression} 

In this section, explicit expressions for each term of $ \sigma^{(1)}(t) $, as listed in Table \ref{table:solution_all_use}, are presented. These expressions, categorized by the perturbations of the Sun, Moon, and Earth's $ J_{2} $, are detailed in \ref{sec:Ana_solution_Sun}, \ref{sec:Ana_solution_Moon}, and \ref{sec:Ana_solution_J2}. Appendix \ref{sec:Ana_solution_lunisolar} provides the  perturbation solutions for jointly solved $ \xi $ and $ \eta $: $ \xi_{c} $, $ \eta_{c} $, $ \xi_{l[c]} $, and $ \eta_{l[c]} $, considering both solar and lunar perturbations. Note that in the subsequent expressions, the orbital elements $ a $, $ \Omega $, $ \lambda $, and $ i $, take the mean value $ \overline{a} $ or the form defined by Eq.~(\ref{eq:sigma_niPJ_msJ2_simplify}). Similarly, the Moon's orbit inclination $ i_{3} $ represents $ \overline{i}_{3} $.

\subsection{Solar perturbation solution $ \sigma_{\text{s}}(t) $}\label{sec:Ana_solution_Sun} 

The secular variation $ \sigma_{\text{s}}^{c}(t) $, long-period variation $ \sigma_{\text{s}}^{l}(t) $, and short-period variation $ \sigma_{\text{s}}^{s}(t) $ within $ \sigma_{\text{s}}(t) $ are presented as follows:

(1) Secular terms with the form $ \sigma_{\text{s}}^{c}(t) $:
\begin{align}
a_{\text{s}}^{c}  = &~  i_{\text{s}}^{c}  = 0, \label{eq:sol_c_sun_inc} \\
\Omega_{\text{s}}^{c} =  &~  n_{\Omega {\text{s}}} \, t , \qquad n_{\Omega {\text{s}}} := - \frac{3}{4} c_\text{s}  \cos i ,\label{eq:sol_c_sun_raan}\\
\lambda_{\text{s}}^{c} =  &~ n_{\lambda \text{s}}  \, t , \qquad n_{\lambda \text{s}} := \frac{1}{8} c_{\text{s}} (1-3 \cos 2 i),
\end{align}
where $ c_\text{s} := \frac{\mu _2}{\overline{n} \,\overline{r}_{2}^3} $.

(2) Long-period terms with the form $ \sigma_{\text{s}}^{l}(t) $:
\begin{align}
 a_{\text{s}}^{l[l]}  = &~ \xi_{\text{s}}^{l[l]}  = \eta_{\text{s}}^{l[l]}  = 0,\\
 i_{\text{s}}^{l[l]}  = & ~ \frac{3}{8} \frac{c_\text{s}}{n_{U_{2}}}\sin i \cos  2 U_{2}
,\\
 \Omega_{\text{s}}^{l[l]} = &  ~ \frac{3}{8} \frac{c_\text{s}  }{ n_{U_{2}}} \cos i \sin 2 U_{2},\\
 \lambda_{\text{s}}^{l[l]} = & ~ \frac{3}{16} \frac{c_\text{s}  }{n_{U_{2}}} (\cos 2 i-3) \sin 2 U_{2},
\end{align}
with
\begin{align} 
& U_{2}(t) := u_{2}(t) - \overline{\Omega}(t) = n_{U_{2}}(t -t_{0}) + U_{2_{0}}, \\
& n_{U_{2}} = n_{2} - n_{\Omega}',\qquad U_{2_{0}} = u_{2_{0}} -  \overline{\Omega}_{0},
\end{align}
where $ n_{\Omega}' $ is the rate of change of $ \overline{\Omega}(t) $ as defined in Eq.~(\ref{eq:sigma_niPJ_msJ2_simplify}). These expressions reveal that solar perturbation induces general long-period variations in satellite orbital elements with an annual period tied to the solar apparent motion $ n_{2} $. The magnitudes of these variations are governed by $ \overline{a} $ and $ \overline{i} $. $ \xi_{\text{s}}^{l[l]} = \eta_{\text{s}}^{l[l]} = 0 $ arises from considering only the leading order $ N = 2 $ within $ P_{N}(\cos \psi_{2} ) $ (refer to Eq.~(\ref{eq:R_s})); when $ N \geq 3 $, both $ \xi $ and $ \eta $ will exhibit periodic variations, as observed in the case of lunar perturbation (see Eqs.~(\ref{eq:xi_ll_moon}) and (\ref{eq:eta_ll_moon})).

There are no special long-period variations in $ a $, $ i $, $ \Omega $, and $ \lambda $,
\begin{align}
\sigma_{\text{s}}^{l[c]} \equiv 0, \quad \text{for}\ \sigma \in \{a, i, \Omega, \lambda \}.
\end{align}
For $ \xi $ and $ \eta $, they exhibit special long-period variations coupled with lunar perturbation, as indicated in Eqs.~(\ref{eq:xi_lc-ms}) and (\ref{eq:eta_lc-ms}).

(3) Short-period terms with the form $ \sigma_{\text{s}}^{s}(t) $:
\begin{align}
 a_{\text{s}}^{s} & =  \sum _{\substack{p=-2 \\ (\Delta p=2)}}^{2} a \times h^{a}_{(2,\, p)} \cos (2 \, \lambda +  p \, U_{2}),\label{eq:sma_s_sun} \\
 \xi_{\text{s}}^{s} & = \sum _{\substack{\kappa = 1 \\ (\Delta \kappa =2)}}^{3}\sum _{\substack{p=-2 \\ (\Delta p=2)}}^{2}  h^{\xi}_{(\kappa, \, p)} \cos (\kappa \, \lambda + p \,  U_{2}),\label{eq:xi_s_sun} \\
 \eta_{\text{s}}^{s} & = \sum _{\substack{\kappa = 1 \\ (\Delta \kappa =2)}}^{3}\sum _{\substack{p=-2 \\ (\Delta p=2)}}^{2}  h^{\eta}_{(\kappa, \, p)} \sin (\kappa \, \lambda + p \,  U_{2}),\label{eq:eta_s_sun}  \\
  i_{\text{s}}^{s} & =  \sum _{\substack{p=-2 \\ (\Delta p=2)}}^{2}  h^{i}_{(2,\, p)} \cos (2 \, \lambda +  p \, U_{2}), \\
  \Omega_{\text{s}}^{s} & =  \sum _{\substack{p=-2 \\ (\Delta p=2)}}^{2}  h^{\Omega}_{(2,\, p)} \sin (2 \, \lambda +  p \, U_{2}), \\  
  \lambda_{\text{s}}^{s} & =  \sum _{\substack{p=-2 \\ (\Delta p=2)}}^{2}  h^{\lambda}_{(2,\, p)} \sin (2 \, \lambda +  p \, U_{2}), 
\end{align}
with
\begin{align} 
h^{\sigma  }_{(\kappa, \, p)}  :=   \frac{  c_\text{s} }{\kappa \, n_{\lambda}'  + p \, n_{U_{2}}}  \times h^{\sigma  }_{[\kappa, \, p]}(i)
\end{align} 
for $ \sigma \in \{a, \xi,\eta, i, \Omega \} $, and
\begin{align}
 h^{\lambda}_{(\kappa, \, p)}  := \frac{  3\, c_{\text{s}} \,n \times  h^{\lambda[a]}_{[\kappa, \, p]}(i)}{( \kappa \,  n_{\lambda}'  + p \, n_{U_{2}} )^{2}}   +  \frac{  c_{\text{s}} \times h^{\lambda[\lambda]}_{[\kappa, \,  p]}(i) }{\kappa \,  n_{\lambda}'  + p \, n_{U_{2}} } ,\label{eq:lambda_s_sun_coeff}
\end{align}
where $ n_{\lambda}' $ denotes the rate of change of $ \overline{\lambda}(t) $ as defined in Eq.~(\ref{eq:sigma_niPJ_msJ2_simplify}). In the specific case of solar perturbation alone, $ n_{\lambda}' = \overline{n} + n_{\lambda \text{s}} $. The terms on the right side of Eq.~(\ref{eq:lambda_s_sun_coeff}) correspond to the integrals of the two terms in Eq.~(\ref{eq:sigma_s1}). The explicit forms of $ h^{\sigma}_{[\kappa, \, p]}(i) $ are given by
\begin{align}
&  h^{a}_{[ 2, 0]}  = \frac{3}{2} \sin ^2 i, \qquad   h^{a}_{[ 2, \pm 2]}  =  \frac{3 }{8} (3 \mp 4 \cos i + \cos 2 i), \nonumber \\
&  h^{ i }_{[ 2, 0]}  = \frac{3  }{8} \sin 2 i , \qquad  h^{ i }_{[ 2, \pm 2]}  =  \frac{3 }{16} (\pm 2 \sin i - \sin 2 i), \nonumber \\
&  h^{ \Omega }_{[ 2, 0]}  = \frac{3}{4} \cos i , \qquad    h^{ \Omega }_{[ 2, \pm 2]}  =   \frac{3}{8} (\pm 1 - \cos i), \nonumber \\
&  h^{  \lambda [a]}_{[ 2, 0]}  =  -\frac{3}{4} \sin ^2 i  , \qquad     h^{  \lambda [a]}_{[ 2, \pm 2]}  = -\frac{3 }{16} (3 \mp 4 \cos i + \cos 2 i),  \nonumber \\
&  h^{  \lambda [\lambda]}_{[ 2, 0]}  = -\frac{3  }{8} (3-\cos 2 i) , \nonumber \\
&    h^{  \lambda [\lambda] }_{[ 2, \pm 2]}  =   - \frac{3 }{16} (5 \mp 6 \cos i + \cos 2 i),   \\
&  h^{ \xi }_{[ 1, 0]}  = \frac{1 }{16} (7-15 \cos 2 i) , \nonumber \\
&  h^{ \xi }_{[ 1, \pm 2]}  = \frac{3 }{32} (7 \mp 12 \cos i + 5 \cos 2 i), \nonumber \\
&  h^{ \xi }_{[ 3, 0]}  = \frac{3}{8} \sin ^2 i  , \qquad  h^{ \xi }_{[ 3, \pm 2]}  = \frac{3 }{32} (3 \mp 4 \cos i + \cos 2 i), \nonumber \\
&  h^{ \eta }_{[ 1, 0]}  = -\frac{ 1}{16 } (11-3 \cos 2 i)  , \nonumber \\
&   h^{ \eta }_{[ 1, \pm 2]}  = -\frac{3 }{32} (11 \mp 12 \cos i +  \cos 2 i), \nonumber \\
&  h^{ \eta }_{[ 3, 0]}  =   \frac{3}{8} \sin ^2 i , \qquad   h^{ \eta }_{[ 3, \pm 2]}  = \frac{3 }{32} (3 \mp 4 \cos i + \cos 2 i). \nonumber 
\end{align}

\subsection{Lunar perturbation solution $ \sigma_{\text{m}}(t) $}  \label{sec:Ana_solution_Moon} 

The secular variation $ \sigma_{\text{m}}^{c}(t) $, special long-period variation $ \sigma_{\text{m}}^{l[c]}(t) $, general long-period variation $ \sigma_{\text{m}}^{l[l]}(t) $, and short-period variation $ \sigma_{\text{m}}^{s}(t) $ within $ \sigma_{\text{m}}(t) $ are shown as follows:

(1) Secular terms with the form $ \sigma_{\text{m}}^{c}(t) $:
\begin{align}
a _{\text{m}}^{c} = &~ i _{\text{m}}^{c} = 0, \label{eq:sol_c_moon_inc}\\
\Omega  _{\text{m}}^{c}=&~ n_{\Omega {\text{m}}} \, t, \label{eq:sol_c_moon_raan}\\
\lambda _{\text{m}}^{c} =&~ n_{\lambda \text{m}} \, t,
\end{align}
where
\begin{align}
 n_{\Omega  {\text{m}}} := & -\frac{3 \, c_ {\text{m}}^{(2)}}{16}  \cos i \, (1+3 \cos 2 i_{3}) -\frac{45 \, c_ {\text{m}}^{(4)} }{32768} \nonumber \\&
\times (9 \cos i+7 \cos 3 i) (9+20 \cos 2 i_{3}+35 \cos 4 i_{3})\nonumber \\&
-\frac{525 \, c_ {\text{m}}^{(6)} }{16777216}(50 \cos i+45 \cos 3 i+33 \cos 5 i) \nonumber \\&
\times(50+105 \cos 2 i_{3} +126 \cos 4 i_{3}+231 \cos 6 i_{3}),\label{eq:sol_c_moon_raan_coeff}
\end{align}
\begin{align}
n_{\lambda \text{m}} := &~ \frac{1}{32} c_{\text{m}}^{(2)} (1-3 \cos 2 i) (1+3 \cos 2 i_{3})  +\frac{9 \, c_{\text{m}}^{(4)}  }{65536} \nonumber \\&
\times (27+40 \cos 2 i-35 \cos 4 i) (9+20 \cos 2 i_{3} \nonumber \\&
 +35 \cos 4 i_{3}) +\frac{75 \, c_{\text{m}}^{(6)}}{33554432}(250+455 \cos 2 i \nonumber \\&
+294 \cos 4 i-231 \cos 6 i) (50+105 \cos 2 i_{3} \nonumber \\&
+126 \cos 4 i_{3}+231 \cos 6 i_{3}),
\end{align}
and $ c_{\text{m}}^{(N)} := \frac{\mu _3}{\overline{n} \, \overline{r}_{2}^3} (\frac{\overline{a}}{\overline{r}_{2}})^{N-2} $.

(2) Special long-period terms with the form $ \sigma_{\text{m}}^{l[c]}(t) $:
\begin{align}
a_{\text{m}}^{l[c]} & = 0, \\
 i_{\text{m}}^{l[c]} &  =  \sum _{\substack{N=2 \\ (\Delta N=2)}}^{ \mathcal{N} } \sum _{q=1}^{N} f^{i (N)}_{(q)} \cos  q\,  \theta_{3},\label{eq:i_lc_moon}  \\
 \Omega_{\text{m}}^{l[c]} &  =  \sum _{\substack{N=2 \\ (\Delta N=2)}}^{ \mathcal{N} } \sum _{q=1}^{N} f^{\Omega (N)}_{(q)} \sin  q\,  \theta_{3}, \label{eq:lambda_lc_moon} \\
  \lambda_{\text{m}}^{l[c]} &  =  \sum _{\substack{N=2 \\ (\Delta N=2)}}^{ \mathcal{N} } \sum _{q=1}^{N} f^{\lambda (N)}_{(q)} \sin  q\,  \theta_{3}, \label{eq:lambda_lc_moon}
\end{align} 
with
\begin{align}
f^{\sigma(N)}_{(q)}  :=  \frac{  c_{\text{m}}^{(N)} }{q \,n_{\theta_{3}}}  \times f^{\sigma(N)}_{[q]}(i), \label{eq:coeff_m_inc_lc}
\end{align} 
and
\begin{align}
& \theta _{3}(t) := \overline{\Omega}(t) - \overline{\Omega} _{3}(t) = n_{\theta _3} (t - t_{0}) + \theta _{3_{0}}, \\
& n_{\theta _3} = n_{\Omega}'-n_{\Omega _3},\qquad \theta _{3_{0}} = \overline{\Omega}_{0} - \Omega_{3_{0}}.
\end{align}
For explicit forms of $ f^{\sigma(N)}_{[q]}(i) $, see Appendix \ref{sec:coeff_moon_lc}.

(3) General long-period terms with the form $ \sigma_{\text{m}}^{l[l]}(t) $:
\begin{align}
 a_{\text{m}}^{l[l]} & = 0,\label{eq:a_ll_moon}
\end{align}
and
\begin{align}
\xi_{\text{m}}^{l[l]} &  =  \sum _{\substack{N=3 \\ (\Delta N=2)}}^{ \mathcal{N} } \sum _{\substack{p > 0 \\ (\Delta p=2)}}^{N} \sum _{q=-N}^{N} f^{\xi (N)}_{(p,\, q)} \cos (p \,  u_{3} + q\,  \theta_{3}),\label{eq:xi_ll_moon} \\
\eta_{\text{m}}^{l[l]} &  =  \sum _{\substack{N=3 \\ (\Delta N=2)}}^{ \mathcal{N} } \sum _{\substack{p > 0 \\ (\Delta p=2)}}^{N} \sum _{\substack{q=-N \\ (q \neq 0)}}^{N} f^{\eta (N)}_{(p,\, q)} \sin (p \,  u_{3} + q\,  \theta_{3}),\label{eq:eta_ll_moon}\\
i_{\text{m}}^{l[l]} &  =  \sum _{\substack{N=2 \\ (\Delta N=2)}}^{ \mathcal{N} } \sum _{\substack{p > 0 \\ (\Delta p=2)}}^{N} \sum _{\substack{q=-N \\ (q \neq 0)}}^{N} f^{i (N)}_{(p,\, q)} \cos (p \,  u_{3} + q\,  \theta_{3}),\label{eq:i_ll_moon}\\
\Omega_{\text{m}}^{l[l]} &  =  \sum _{\substack{N=2 \\ (\Delta N=2)}}^{ \mathcal{N} } \sum _{\substack{p > 0 \\ (\Delta p=2)}}^{N} \sum _{q=-N}^{N} f^{\Omega (N)}_{(p,\, q)} \sin (p \,  u_{3} + q\,  \theta_{3}),\label{eq:Omega_ll_moon}\\
\lambda_{\text{m}}^{l[l]} &  =  \sum _{\substack{N=2 \\ (\Delta N=2)}}^{ \mathcal{N} } \sum _{\substack{p > 0 \\ (\Delta p=2)}}^{N} \sum _{q=-N}^{N} f^{\lambda (N)}_{(p,\, q)} \sin (p \,  u_{3} + q\,  \theta_{3}),\label{eq:lambda_ll_moon}
\end{align}
with
\begin{align}
f^{\sigma(N)}_{(p,\, q)}  := \frac{  c_{\text{m}}^{(N)} }{p \, n_{3} + q \,n_{\theta_{3}}}  \times f^{\sigma(N)}_{[p,\, q]}(i). \label{eq:coeff_m_inc_ll} 
\end{align} 
For explicit forms of $ f^{\sigma(N)}_{[p,\, q]}(i) $, see Appendix \ref{sec:coeff_moon_ll}.

(4) Short-period terms with the form $ \sigma_{\text{m}}^{s}(t) $:
\begin{align}
a_{\text{m}}^{s} & = ~ \sum _{N=2}^{ \mathcal{N} } \sum _{\substack{\kappa > 0 \\ (\Delta \kappa =2)}}^{N}\sum _{\substack{p=-N \\ (\Delta p=2)}}^{N} \sum _{q=-N}^{N} a \, f^{a(N)}_{(\kappa,\, p,\, q)} \cos \Lambda,  \label{eq:as_m} \\
\xi_{\text{m}}^{s} & = \sum _{N=2}^{ \mathcal{N} } \sum _{\substack{\kappa > 0 \\ (\Delta \kappa =2)}}^{N+1}\sum _{\substack{p=-N \\ (\Delta p=2)}}^{N} \sum _{q=-N}^{N} f^{\xi(N)}_{(\kappa,\, p,\, q)} \cos \Lambda,\label{eq:xi_s_moon}\\
\eta_{\text{m}}^{s} & =  \sum _{N=2}^{ \mathcal{N} } \sum _{\substack{\kappa > 0 \\ (\Delta \kappa =2)}}^{N+1}\sum _{\substack{p=-N \\ (\Delta p=2)}}^{N} \sum _{q=-N}^{N} f^{\eta(N)}_{(\kappa,\, p,\, q)} \sin \Lambda,\label{eq:eta_s_moon}\\
i_{\text{m}}^{s} & = \sum _{N=2}^{ \mathcal{N} } \sum _{\substack{\kappa > 0 \\ (\Delta \kappa =2)}}^{N}\sum _{\substack{p=-N \\ (\Delta p=2)}}^{N} \sum _{q=-N}^{N} f^{i(N)}_{(\kappa,\, p,\, q)} \cos \Lambda,\label{eq:i_s_moon}\\
\Omega_{\text{m}}^{s} & =  \sum _{N=2}^{ \mathcal{N} } \sum _{\substack{\kappa > 0 \\ (\Delta \kappa =2)}}^{N}\sum _{\substack{p=-N \\ (\Delta p=2)}}^{N} \sum _{q=-N}^{N} f^{\Omega(N)}_{(\kappa,\, p,\, q)} \sin \Lambda,\label{eq:Omega_s_moon}\\
\lambda_{\text{m}}^{s} & =  \sum _{N=2}^{ \mathcal{N} } \sum _{\substack{\kappa > 0 \\ (\Delta \kappa =2)}}^{N}\sum _{\substack{p=-N \\ (\Delta p=2)}}^{N} \sum _{q=-N}^{N} f^{\lambda(N)}_{(\kappa,\, p,\, q)} \sin \Lambda,\label{eq:lambda_s_moon}
\end{align}
with $ \Lambda := \kappa \, \lambda + p \,  u_{3} + q\,  \theta_{3} $,
\begin{align}
f^{\sigma(N)}_{(\kappa,\, p,\, q)}  :=   \frac{  c_{\text{m}}^{(N)} }{\kappa \,  n_{\lambda}' + p \, n_{3} + q \,n_{\theta_{3}}}  \times f^{\sigma(N)}_{[\kappa,\, p,\, q]}(i)  \label{eq:coeff_m_sma_s} 
\end{align}
for $ \sigma \in \{a, \xi,\eta, i, \Omega \} $, and
\begin{align}
f^{\lambda(N)}_{(\kappa,\, p,\, q)} :=&~  \frac{  3\, c_{\text{m}}^{(N)}\,n }{(\kappa \, n_{\lambda}' + p \, n_{3} + q \,n_{\theta_{3}})^{2}}  \times f^{\lambda(N)[a]}_{[\kappa,\, p,\, q]}(i) \nonumber \\
& + \frac{  c_{\text{m}}^{(N)} }{\kappa \,  n_{\lambda}' + p \, n_{3} + q \,n_{\theta_{3}}}  \times f^{\lambda(N)[\lambda]}_{[\kappa,\, p,\, q]}(i) . \label{eq:lambda_s_moon_coeff}
\end{align} 
For explicit forms of $ f^{\sigma(N)}_{[\kappa,\, p,\, q]}(i) $, see Appendix \ref{sec:coeff_moon_s}.

\subsection{Earth's $ J_{2} $ perturbation solution $ \sigma _{\! J_{2}}(t) $}\label{sec:Ana_solution_J2} 

The Earth's $J_{2}$ perturbation solution, $ \sigma _{\! J_{2}\text{(eq)}} $ in the geocentric equatorial coordinate system, can be derived straightforwardly from Eqs.~(\ref{eq:R_J2}) and (\ref{eq:sigma_c1})-(\ref{eq:sigma_s1}). For consistency with the solar and lunar perturbation solutions, the ecliptic representation $ \sigma _{\! J_{2}} $ is essential. Given that the $ J_{2} $ perturbation is significantly smaller than lunisolar perturbations, this study focuses on the secular components, including $ i_{J2}^{c} $, $ \Omega_{J2}^{c} $, and $ \lambda_{J2}^{c} $. Employing spherical trigonometry \cite{Liu2000,Capderou2014} and variational method, we have
\begin{align}
i _ {J_{2}}^{c}&= \sin \Omega  \sin \epsilon \, \Omega _ {J_ {2}\text{(eq)}}^{c} = n_{i _{J_{2}}}  t,\label{eq:i_J2_c}\\
\Omega _ {J_{2}}^{c}&=(\cos \epsilon + \cot i \cos \Omega  \sin \epsilon  )\, \Omega _ {J_ {2}\text{(eq)}}^{c} = n_{\Omega _{J_{2}}}  t,\label{eq:Omega_J2_c}\\
\lambda _ {J_{2}}^{c}&=\lambda _{J_{2}\text{(eq)}}^{c}-\frac{ \cos \Omega  }{\sin i}\sin \epsilon 
 \, \Omega _ {J_ {2}\text{(eq)}}^{c} = n_{\lambda _{J_{2}}} \, t,\label{eq:lambda_J2_c}
\end{align}
with
\begin{align}
n_{i _{J_{2}}} & := - \frac{3}{2} c_{J_{2}} \sin \Omega  \sin \epsilon \cos i_\text{eq} ,\\
n_{\Omega _{J_{2}}} & :=  - \frac{3}{2} c_{J_{2}} (\cos \epsilon + \cot i \cos \Omega  \sin \epsilon  ) \cos i_\text{eq} ,\\
n_{\lambda _{J_{2}}} & :=  \frac{3}{2} c_{\small{J_2}}  (1 + 2 \cos 2 i_\text{eq}) + \frac{3}{2} c_{J_{2}} \cos i_\text{eq} \frac{ \cos \Omega  }{\sin i}\sin \epsilon,
\end{align}
where $ \epsilon = 23.439\,291^{\circ} $ is the obliquity of the ecliptic, $ c_{\small{J_2}} := \frac{ \mu J_2 R_{\text{e}}^2}{  \overline{a}^5 \, \overline{n} } $, and 
\begin{align}
\cos i_\text{eq}=\cos i \cos \epsilon -\sin i \cos \Omega  \sin \epsilon.\label{eq:cos_iEq-text}
\end{align}

\subsection{Expressions for $ \xi_{c} $, $ \eta_{c} $, $ \xi_{l[c]} $, and $ \eta_{l[c]} $}\label{sec:Ana_solution_lunisolar} 

As pointed out in Appendix \ref{sec:Ana_solution_DerivVerif12}, deriving the ``secular" perturbation solutions $ \xi^{c}_{\text{new}}(t) $ and $ \eta^{c}_{\text{new}}(t) $,
\begin{align}
\xi^{c}_{\text{new}}(t) & := \xi_{c}(t)+\xi_{l[c]}(t), \label{eq:xi_c_new} \\
\eta^{c}_{\text{new}}(t) & := \eta_{c}(t)+\eta_{l[c]}(t), \label{eq:eta_c_new}
\end{align}
involves solving the oscillation equations:
\begin{align}
\frac{\mathrm{d} \xi^{c}_{\text{new}}(t)}{\mathrm{d}t}  & =    h_{\xi} \, \eta^{c}_{\text{new}}(t) + f_{1\xi}^{l[c]}(t), \label{eq:dxidt_ms-c} \\
\frac{\mathrm{d}\eta^{c}_{\text{new}}(t)}{\mathrm{d}t} & =  h_{\eta}  \, \xi^{c}_{\text{new}}(t) + f_{1\eta}^{l[c]}(t).
\label{eq:detadt_ms-c}
\end{align}
Due to the minor eccentricity variations ($ \sim $$10^{-6}$) induced by the $ J_{2} $ perturbation, only the effects of the lunisolar perturbations are taken into account, resulting in expressions for $ h_{\xi} $ and $ h_{\eta} $ given by 
\begin{align}
h_{\xi} \simeq  h_\text{s}^{\xi} +  h_\text{m}^{\xi},\qquad h_{\eta} \simeq h_\text{s}^{\eta} + h_\text{m}^{\eta},\label{eq:xi&eta_Cf}
\end{align}
with
\begin{align}
h_\text{s}^{\xi}=\frac{3}{8} c_\text{s} (1-5 \cos 2 i),\qquad h_\text{s}^{\eta}=\frac{3}{2}c_\text{s},\label{eq:xi&eta_Cf-s}
\end{align}
and
\begin{align}
h_\text{m}^{\xi}=&~ \frac{3 \, c_ {\text{m}}^{(2)}}{32}  (1-5 \cos 2 i) (1+ 3 \cos 2 i_ 3) -\frac{45 \, c_ {\text{m}}^{(4)}}{65536} (3 \nonumber \\ 
& +12 \cos 2 i+49 \cos 4 i) (9+20 \cos 2 i_ 3+35 \cos 4 i_ 3)  \nonumber \\ 
& -\frac{525 \, c_ {\text{m}}^{(6)} }{33554432}(30 + 71 \cos 2 i+114 \cos 4 i+297 \cos 6 i) \nonumber \\ 
& \times (50 + 105 \cos 2 i_ 3 +126 \cos 4 i_ 3+231 \cos 6 i_ 3),\nonumber \\
h_\text{m}^{\eta}=&~ \frac{3 \, c_ {\text{m}}^{(2)}}{8}  (1+3 \cos 2 i_ 3) \nonumber \\
& +\frac{45 \, c_ {\text{m}}^{(4)}}{8192}(3+5 \cos 2 i) (9+20 \cos 2 i_ 3+35 \cos 4 i_ 3) \nonumber \\
& +\frac{525 \, c_ {\text{m}}^{(6)}}{4194304}(15 + 28 \cos 2 i+21 \cos 4 i) \nonumber \\ 
& \times (50 + 105 \cos 2 i_ 3+126 \cos 4 i_ 3  +231 \cos 6 i_ 3).\label{eq:eta_Cf-m}
\end{align} 
Moreover, the expressions for $ f_{1\xi}^{l[c]}(t) $ and $ f_{1\eta}^{l[c]}(t) $ are
\begin{align}
f_{1\xi}^{l[c]} & = \sum _{q=-5}^{5} - f^{\xi}_{( q)} \sin (\Delta u_{3} + q\,  \theta_{3}),\\
f_{1\eta}^{l[c]} & =  \sum _{\substack{q=-5 \\ (q \neq 0)}}^{5} f^{\eta}_{( q)} \cos (\Delta u_{3} + q\, \theta_{3}),
\end{align}
with
\begin{align}
f^{\xi}_{( q)} & :=  c_{\text{m}}^{(3)}(4 \, r_3^{\varepsilon }+u_3^{A}) f^{\xi(3)}_{[q]}(i)  + c_{\text{m}}^{(5)}(6 \, r_3^{\varepsilon }+u_3^{A}) f^{\xi(5)}_{[q]}(i) ,\label{eq:f_xi(q)}\\
f^{\eta}_{( q)} & := c_{\text{m}}^{(3)}(4 \, r_3^{\varepsilon }+u_3^{A})  f^{\eta(3)}_{[q]}(i) + c_{\text{m}}^{(5)}(6 \, r_3^{\varepsilon }+u_3^{A}) f^{\eta(5)}_{[q]}(i),\label{eq:f_eta(q)}
\end{align}
and
\begin{align}
& \Delta u_{3}(t) := u_{3}(t) - \overline{M}_{3}(t) = \Delta n_{3} \, (t-t_{0}) + \Delta u_{3_{0}}, \label{eq:Delta_u3} \\
& \Delta n_{3} = n_{3} - n_{\!M_3},\qquad \Delta u_{3_{0}} =   u_{3_{0}} - M_{3_{0}},
\end{align}
where $ r_3^{\varepsilon } := \frac{r_3^{A}}{\overline{r}_{3}} $, and $ f^{\eta}_{(0)} = 0 $. Further, solving Eqs.~(\ref{eq:dxidt_ms-c}) and (\ref{eq:detadt_ms-c}) yields
\begin{align}
& \xi_{c}(t) =  c_{1} \cosh (\sqrt{h_{\xi}} \sqrt{h_{\eta}}  t ) + c_{2} \frac{\sqrt{h_{\xi}} }{\sqrt{h_{\eta}}}\sinh (\sqrt{h_{\xi}} \sqrt{h_{\eta}} t ),\label{eq:xi_c-ms}\\
& \eta_{c}(t) = c_{2} \cosh (\sqrt{h_{\xi}} \sqrt{h_{\eta}}  t )+ c_{1} \frac{\sqrt{h_{\eta}} }{\sqrt{h_{\xi}}} \sinh (\sqrt{h_{\xi}} \sqrt{h_{\eta}} t ),\label{eq:eta_c-ms} \\
& \xi_{l[c]}(t) =  \sum _{q=-5}^{5} \frac{ f^{\xi}_{( q)} (\Delta n_{3} + q \,  n_{\theta_{3}}) -h_{\xi} 
f^{\eta}_{( q)} }{ (\Delta n_{3} + q \,  n_{\theta_{3}})^{2} +h_{\xi} h_{\eta} } \cos \Gamma,\label{eq:xi_lc-ms}\\
& \eta_{l[c]}(t) =  \sum _{q=-5}^{5} \frac{ f^{\eta}_{( q)}  (\Delta n_{3} + q \,  n_{\theta_{3}}) +h_{\eta} f^{\xi}_{( q)} }{ (\Delta n_{3} + q \,  n_{\theta_{3}})^{2} +h_{\xi} h_{\eta} }  \sin \Gamma,\label{eq:eta_lc-ms}
\end{align}
where  
\begin{align}
c_{1} =  \overline{\xi}_{\!0} - \xi_{l[c]}(t_{0}),\qquad  c_{2} = &~ \overline{\eta}_{\!0} -  \eta_{l[c]}(t_{0}),\label{eq:c1c2}
\end{align}
and $ \Gamma := \Delta u_{3}+ q\,  \theta_{3} $.
Equations (\ref{eq:xi_lc-ms}) and (\ref{eq:eta_lc-ms}) indicate that solar perturbation alone induces no special long-period  variations in $ \xi $ and $ \eta $, since the terms $ f^{\xi}_{( q)} $ and $ f^{\eta}_{( q)} $ are exclusively associated with lunar perturbation. Additionally, it is worth noting that introducing $ \sigma _{l[c]}(t) $ in the reference solution (\ref{eq:sigmaPJt_NiPJ}), as well as in Eqs.~(\ref{eq:xi_c_new}) and (\ref{eq:eta_c_new}), is crucial. Without this term, a significant increase in analytical solution errors would occur, leading to the disappearance of terms related to $ \xi_{l[c]}(t_{0}) $ and $  \eta_{l[c]}(t_{0}) $ in Eq.~(\ref{eq:c1c2}), as well as the terms associated with $ h_{\xi} $ and $ h_{\eta} $ in Eqs.~(\ref{eq:xi_lc-ms}) and (\ref{eq:eta_lc-ms}).

\section{Explicit forms of inclination functions}
\label{sec:app_coeff_moon}

The explicit forms of inclination functions within the lunar perturbation solution are shown below. These encompass $ f^{\sigma(N)}_{[q]}(i) $ in Eqs.~(\ref{eq:coeff_m_inc_lc}), (\ref{eq:f_xi(q)}), and (\ref{eq:f_eta(q)})  for the special long-period terms, $ f^{\sigma(N)}_{[p,\, q]}(i) $ in Eq.~(\ref{eq:coeff_m_inc_ll}) for the general long-period terms, and $ f^{\sigma(N)}_{[\kappa,\, p,\, q]}(i) $ in Eqs.~(\ref{eq:coeff_m_sma_s}) and (\ref{eq:lambda_s_moon_coeff}) for the short-period terms. For brevity, only the leading-order inclination functions with $ N = 2 $ or $ N = 3 $ are presented (cf.~Table \ref{table:Coeff_structure_m_sol}). Inclination functions for other orders can be derived using the methods outlined in Appendix \ref{sec:Ana_solution_DerivVerif12}. Note that $ i $ and $ i_{3} $ in these functions represent mean values $ \overline{i} $ and $ \overline{i}_{3} $, respectively.

\begin{table}[ht]   
  \caption{Relationship between the inclination functions and the Legendre polynomial degree $ N $.}   
    \begin{ruledtabular}
    \begin{tabular}{lccc}
$ \sigma $   &  $ f^{\sigma (N)}_{[q]}(i) $  &   $ f^{\sigma (N)}_{[p,\, q]}(i) $  & $ f^{\sigma(N)}_{[\kappa,\, p,\, q]}(i) $  \\
        \hline
$ a $  &  $ - $   &  $ - $  &   $ N = 2,3,4,\cdots $  \\
$ \xi $, $ \eta $   & $ N = 3,5,\cdots $   & $ N = 3,5,\cdots $  &  $ N = 2,3,4,\cdots $  \\
$ i $, $ \Omega $, $ \lambda  $  &  $ N = 2,4,\cdots $     &  $ N = 2,4,\cdots $    &  $ N = 2,3,4,\cdots $    \\
\end{tabular}
\end{ruledtabular}
    \label{table:Coeff_structure_m_sol}
\end{table}

\subsection{Inclination functions for special long-period terms}
\label{sec:coeff_moon_lc}

The inclination functions $  f^{\sigma (N)}_{[q]}(i) $, associated with the special long-period terms $ \sigma_{\text{m}}^{l[c]}(t) $ in Eqs.~(\ref{eq:xi_lc-ms})-(\ref{eq:eta_lc-ms}) and (\ref{eq:i_lc_moon})-(\ref{eq:lambda_lc_moon}), are listed below in terms of the orbital elements $ \xi $, $ \eta $, $ i $, $ \Omega $, and $ \lambda $:

(1) Inclination functions with the form $  f^{\xi (3)}_{[q]}(i) $:
\begin{align}
&\displaystyle f^{\xi(3)}_{[ \raisebox{0mm}{-}3]} = \frac{225}{32768}  \frac{(\cos  i-\cos  3 i) (3-4 \cos  2  i_{3}+\cos  4  i_{3})}{1-\cos   i_{3}} , \nonumber \\
&\displaystyle f^{\xi(3)}_{[ 0]} = \frac{45}{4096}  (\sin  i+5 \sin  3 i) (\sin
     i_{3}+5 \sin  3  i_{3}) , \nonumber \\
&\displaystyle f^{\xi(3)}_{[ \pm 1]} = -\frac{15}{16384}  (\cos  i+15 \cos  3 i) (\pm 6-\cos
     i_{3}   \nonumber\\
&\displaystyle\phantom{f^{\xi(3)}_{[ \pm 1]} =}   \pm 10 \cos  2  i_{3} -15 \cos  3  i_{3}) , \nonumber \\
&\displaystyle f^{\xi(3)}_{[ \pm 2]} = \frac{75}{8192}  (\sin  i-3 \sin  3 i) (\sin
     i_{3} \pm 4 \sin  2  i_{3} \nonumber \\
&\displaystyle\phantom{f^{\xi(3)}_{[ \pm 2]}=}      -3 \sin  3  i_{3}) , \nonumber \\
&\displaystyle f^{\xi(3)}_{[ 3]} = -\frac{225}{16384}  (\cos  i-\cos  3 i) (2-\cos
     i_{3}-2 \cos  2  i_{3}  \nonumber\\
&\displaystyle\phantom{f^{\xi(3)}_{[ 3]} =}     +\cos  3  i_{3}).
\end{align}

(2) Inclination functions with the form $  f^{\eta (3)}_{[q]}(i) $:
\begin{align}
 &\displaystyle f^{\eta(3)}_{[ \raisebox{0mm}{-}3]} = \frac{225}{16384}  \frac{(1-\cos  2 i) (3-4 \cos  2  i_{3}+\cos  4  i_{3})}{1-\cos   i_{3}} , \nonumber \\
  &\displaystyle f^{\eta(3)}_{[ \pm 1]} = \frac{15}{8192}  (3+5 \cos  2 i) (6 \mp \cos   i_{3}+10
   \cos  2  i_{3} \nonumber \\
&\displaystyle\phantom{f^{\eta(3)}_{[ \pm 1]} = }    \mp 15 \cos  3  i_{3}) ,  \\
  &\displaystyle f^{\eta(3)}_{[ \pm 2]} = \frac{75}{2048}  \sin  2 i ( \pm \sin   i_{3}+4 \sin  2  i_{3} \mp 3 \sin  3  i_{3}) , \nonumber \\
  &\displaystyle f^{\eta(3)}_{[ 3]} = \frac{225}{8192}  (1-\cos  2 i) (2-\cos   i_{3}-2 \cos
    2  i_{3}+\cos  3  i_{3}). \nonumber
\end{align}

(3) Inclination functions with the form $  f^{i (2)}_{[q]}(i) $:
\begin{align}
& f^{i(2)}_{[1]} = -\frac{3}{8}  \cos  i \sin  2 i_{3} , \nonumber \\
& f^{i(2)}_{[2]} = -\frac{3}{16}  (1-\cos  2 i_{3}) \sin  i. 
\end{align}

(4) Inclination functions with the form $ f^{\Omega(2)}_{[q]}(i) $:
\begin{align}
& f^{\Omega(2)}_{[1]} = \frac{3}{8}  \cos  2 i \csc  i \sin  2 i_{3} , \nonumber \\
& f^{\Omega(2)}_{[2]} = \frac{3}{16}  \cos  i (1-\cos  2 i_{3}). 
\end{align}

(5) Inclination functions with the form $  f^{\lambda(2)}_{[q]}(i) $:
\begin{align}
& f^{\lambda(2)}_{[1]} = -\frac{3}{16}  (3 \cos  i-\cos  3 i) \csc  i \sin  2 i_{3} ,  \nonumber \\
& f^{\lambda(2)}_{[2]} = -\frac{3}{32}  (3-\cos  2 i) (1-\cos  2 i_{3}).
\end{align}

\subsection{Inclination functions for general long-period terms} 
\label{sec:coeff_moon_ll}

The inclination functions $  f^{\sigma (N)}_{[p,\, q]}(i) $ for the general long-period terms $ \sigma_{\text{m}}^{l[l]}(t) $ in Eqs.~(\ref{eq:xi_ll_moon})-(\ref{eq:lambda_ll_moon}) are listed in the order of $ \xi $, $ \eta $, $ i $, $ \Omega $, and $ \lambda $, as follows:

(1) Inclination functions with the form $  f^{\xi (3)}_{[p,\, q]}(i) $:
\begin{align}
 &\displaystyle f^{\xi(3)}_{[1,\raisebox{0mm}{-}3]} = -\frac{225}{16384}  \frac{(\cos  i-\cos  3 i) (3-4 \cos
    2  i_{3}+\cos  4  i_{3})}{1-\cos   i_{3}} , \nonumber \\
     &\displaystyle f^{\xi(3)}_{[1,0]} = -\frac{45}{2048}  (\sin  i+5 \sin  3 i) (\sin
     i_{3}+5 \sin  3  i_{3}) , \nonumber \\
 &\displaystyle f^{\xi(3)}_{[1, \pm 1]} = \frac{15}{8192}  (\cos  i+15 \cos  3 i) (\pm 6-\cos
     i_{3} \pm 10 \cos  2  i_{3} \nonumber \\
     &\displaystyle\phantom{f^{\xi(3)}_{[1, \pm 1]} =} -15 \cos  3  i_{3}) , \nonumber \\
 &\displaystyle f^{\xi(3)}_{[1, \pm 2]} = -\frac{75}{4096}  (\sin  i-3 \sin  3 i) (\sin
     i_{3} \pm 4 \sin  2  i_{3} \nonumber \\
     &\displaystyle\phantom{f^{\xi(3)}_{[1, \pm 2]} =} -3 \sin  3  i_{3}) , \nonumber \\
 &\displaystyle f^{\xi(3)}_{[1, 3]} = \frac{225}{8192}  (\cos  i-\cos  3 i) (2-\cos
     i_{3}-2 \cos  2  i_{3} \nonumber \\
     &\displaystyle\phantom{f^{\xi(3)}_{[1, 3]} =} +\cos  3  i_{3}) ,\nonumber  \\
 &\displaystyle f^{\xi(3)}_{[3,\raisebox{0mm}{-}2]} = \frac{75(3 \sin  3 i - \sin  i) (10 \sin
     i_{3}-5 \sin  3  i_{3}+\sin  5  i_{3})}{8192 (3-4 \cos   i_{3}+\cos  2  i_{3})} , \nonumber \\
 &\displaystyle f^{\xi(3)}_{[3,0]} = -\frac{75}{2048}  (\sin  i+5 \sin  3 i) (3 \sin
     i_{3}-\sin  3  i_{3}) , \nonumber \\
 &\displaystyle f^{\xi(3)}_{[3,\pm 1]} = \frac{75}{8192}  (\cos  i+15 \cos  3 i) ( \pm 2-\cos
     i_{3} \mp 2 \cos  2  i_{3} \nonumber \\
 &\displaystyle\phantom{f^{\xi(3)}_{[3,\pm 1]} = } +\cos  3  i_{3}) , \nonumber \\
 &\displaystyle f^{\xi(3)}_{[3,2]} = \frac{75}{4096}  (3 \sin  3 i -\sin  i) (5 \sin
     i_{3}-4 \sin  2  i_{3}+\sin  3  i_{3}) , \nonumber \\
 &\displaystyle f^{\xi(3)}_{[3,\pm 3]} = \frac{75}{8192}  (\cos  i-\cos  3 i) (\pm 10-15 \cos
     i_{3} \pm 6 \cos  2  i_{3} \nonumber \\
 &\displaystyle\phantom{f^{\xi(3)}_{[3,\pm 3]} =} -\cos  3  i_{3}).  
\end{align}

(2) Inclination functions with the form $  f^{\eta (3)}_{[p,\, q]}(i) $:
\begin{align}
 &\displaystyle f^{\eta(3)}_{[1,\raisebox{0mm}{-}3]} = -\frac{225}{8192}  \frac{(1-\cos  2 i) (3-4 \cos  2  i_{3}+\cos  4  i_{3})}{1-\cos   i_{3}} , \nonumber \\
 &\displaystyle f^{\eta(3)}_{[1, \pm 1]} = -\frac{15}{4096}  (3+5 \cos  2 i) (6 \mp \cos   i_{3}+10
   \cos  2  i_{3} \nonumber \\
 &\displaystyle\phantom{f^{\eta(3)}_{[1, \pm 1]} =} \mp 15 \cos  3  i_{3}) , \nonumber \\
 &\displaystyle f^{\eta(3)}_{[1, \pm 2]} = -\frac{75}{1024}  \sin  2 i (\pm \sin   i_{3}+4 \sin  2  i_{3} \mp 3 \sin  3  i_{3}) , \nonumber \\
 &\displaystyle f^{\eta(3)}_{[1,3]} = \frac{225}{4096}  (\cos  2 i -1) (2-\cos   i_{3}-2
   \cos  2  i_{3}+\cos  3  i_{3}) , \nonumber \\
 &\displaystyle f^{\eta(3)}_{[3, \pm 1]} = -\frac{75}{4096}  (3+5 \cos  2 i) (2 \mp \cos   i_{3}-2
   \cos  2  i_{3} \nonumber \\
 &\displaystyle\phantom{f^{\eta(3)}_{[3, \pm 1]} =} \pm \cos  3  i_{3}) , \nonumber \\
 &\displaystyle f^{\eta(3)}_{[3, \pm 2]} = -\frac{75}{1024}  \sin  2 i (\pm 5 \sin   i_{3}-4 \sin  2  i_{3} \pm \sin  3  i_{3}) , \nonumber \\
 &\displaystyle f^{\eta(3)}_{[3, \pm 3]} = -\frac{75}{4096}  (1-\cos  2 i) (10 \mp 15 \cos   i_{3}+6
   \cos  2  i_{3} \nonumber \\
 &\displaystyle\phantom{f^{\eta(3)}_{[3, \pm 3]} =}  \mp \cos  3  i_{3}).  
\end{align}

(3) Inclination functions with the form $  f^{i(2)}_{[p,\, q]}(i) $:
\begin{align}
 & f^{i(2)}_{[ 2, \pm 1]} = -\frac{3}{16}  \cos   i (2 \sin  i_{3} \mp \sin  2 i_{3}) ,\nonumber  \\
 & f^{i(2)}_{[ 2, \pm 2]} = -\frac{3}{32}  (\pm 3-4 \cos  i_{3} \pm \cos  2 i_{3}) \sin   i. 
\end{align}

(4) Inclination functions with the form $  f^{\Omega (2)}_{[p,\, q]}(i) $:
\begin{align}
 & f^{\Omega(2)}_{[ 2, 0]} = -\frac{9}{16}  \cos   i (1-\cos  2 i_{3}) , \nonumber \\
 & f^{\Omega(2)}_{[ 2, \pm 1]} = \frac{3}{16}  \cos  2  i \csc   i (\pm 2 \sin  i_{3}-\sin  2 i_{3}) , \nonumber \\
 & f^{\Omega(2)}_{[ 2, \pm 2]} = \frac{3}{32}  \cos   i (3 \mp 4 \cos  i_{3}+\cos  2 i_{3}). 
\end{align}

(5) Inclination functions with the form $  f^{\lambda (2)}_{[p,\, q]}(i) $:
\begin{align}
 & f^{\lambda(2)}_{[ 2, 0]} = \frac{3}{32}  (1-3 \cos  2  i) (1-\cos  2 i_{3}) , \nonumber \\
 & f^{\lambda(2)}_{[ 2, \pm 1]} = -\frac{3}{32}  (3 \cos   i-\cos  3  i) \csc   i (\pm 2 \sin  i_{3}-\sin  2 i_{3}) , \nonumber \\
 & f^{\lambda(2)}_{[ 2, \pm 2]} = -\frac{3}{64}  (3-\cos  2  i) (3 \mp 4 \cos  i_{3}+\cos  2 i_{3}). 
\end{align}

\subsection{Inclination functions for short-period terms} 
\label{sec:coeff_moon_s}

The inclination functions $  f^{\sigma(N)}_{[\kappa,\, p,\, q]}(i) $ for the short-period terms $ \sigma_{\text{m}}^{s}(t) $ in Eqs.~(\ref{eq:as_m})-(\ref{eq:lambda_s_moon}) are listed below in the order of $ a $, $ \xi $, $ \eta $, $ i $, $ \Omega $, and $ \lambda $:

(1) Inclination functions with the form $  f^{a (2)}_{[\kappa,\, p,\, q]}(i) $:
\begin{align}
&  f^{a(2)}_{[  2,\raisebox{0mm}{-} 2, 0]}  = \frac{9}{32}  (1-\cos  2  i) (1-\cos  2 i_{3}) , \nonumber \\
&  f^{a(2)}_{[  2,\raisebox{0mm}{-} 2, \pm 1]}  = \frac{3}{16}  (2 \sin   i \pm \sin  2  i) (2 \sin  i_{3} \pm \sin  2 i_{3}) , \nonumber \\
&  f^{a(2)}_{[  2,\raisebox{0mm}{-} 2, \pm 2]}  = \frac{3}{64}  (3 \pm 4 \cos   i+\cos  2  i) (3 \pm 4 \cos  i_{3}+\cos  2 i_{3}) , \nonumber \\ 
&  f^{a(2)}_{[  2, 0, 0]}  = \frac{3}{16}  (1-\cos  2  i) (1+3 \cos  2 i_{3}) , \nonumber \\
&  f^{a(2)}_{[  2, 0, \pm 1]}  = -\frac{3}{8}  (\pm 2 \sin   i+\sin  2  i) \sin  2 i_{3} ,  \\
&  f^{a(2)}_{[  2, 0, \pm 2]}  = \frac{3}{32}  (3 \pm 4 \cos   i+\cos  2  i) (1-\cos  2 i_{3}) , \nonumber \\
&  f^{a(2)}_{[  2, 2, 0]}  =  f^{a(2)}_{[  2,\raisebox{0mm}{-} 2, 0]} , \nonumber \\
&  f^{a(2)}_{[  2, 2, \pm 1]}  = -\frac{3}{16}  (2 \sin   i \pm \sin  2  i) (2 \sin  i_{3} \mp \sin  2 i_{3}) , \nonumber \\
&  f^{a(2)}_{[  2, 2, \pm 2]}  = \frac{3}{64}  (3 \pm 4 \cos   i+\cos  2  i) (3 \mp 4 \cos  i_{3}+\cos  2 i_{3}). \nonumber 
\end{align}

(2) Inclination functions with the form $  f^{\xi (2)}_{[\kappa,\, p,\, q]}(i) $:
\begin{align}
&\displaystyle f^{\xi(2)}_{[ 1,\raisebox{0mm}{-} 2, 0]}  =  \frac{3}{128}  (7-15 \cos  2  i) (1-\cos  2 i_{3}) , \nonumber  \\
&\displaystyle f^{\xi(2)}_{[ 1,\raisebox{0mm}{-} 2, \pm 1]}  =  \frac{3}{64}  (6 \sin   i \pm 5 \sin  2  i) (2 \sin  i_{3} \pm \sin  2 i_{3}) , \nonumber \\
&\displaystyle f^{\xi(2)}_{[ 1,\raisebox{0mm}{-} 2, \pm 2]}  =  \frac{3}{256}  (7 \pm 12 \cos   i+5 \cos  2  i) (3 \pm 4 \cos  i_{3} \nonumber \\
&\displaystyle\phantom{f^{\xi(2)}_{[ 1,\raisebox{0mm}{-} 2, \pm 2]}  =}  +\cos  2 i_{3}) , \nonumber \\
&\displaystyle f^{\xi(2)}_{[ 1, 0, 0]}  =  \frac{1}{64}  (7-15 \cos  2  i) (1+3 \cos  2 i_{3}) , \nonumber \\
&\displaystyle f^{\xi(2)}_{[ 1, 0, \pm 1]}  =  -\frac{3}{32}  (\pm 6 \sin   i+5 \sin  2  i) \sin  2 i_{3} , \nonumber \\
&\displaystyle f^{\xi(2)}_{[ 1, 0, \pm 2]}  =  \frac{3}{128}  (7 \pm 12 \cos   i+5 \cos  2  i) (1-\cos  2 i_{3}) , \nonumber \\
&\displaystyle f^{\xi(2)}_{[ 1, 2, 0]}  =  f^{\xi(2)}_{[ 1,\raisebox{0mm}{-} 2, 0]} , \nonumber \\
&\displaystyle f^{\xi(2)}_{[ 1, 2, \pm 1]}  =  -\frac{3}{64}  (6 \sin   i \pm 5 \sin  2  i) (2 \sin  i_{3} \mp \sin  2 i_{3}) , \nonumber \\
&\displaystyle f^{\xi(2)}_{[ 1, 2, \pm 2]}  =  \frac{3}{256}  (7 \pm 12 \cos   i+5 \cos  2  i) (3 \mp 4 \cos  i_{3} \nonumber \\
&\displaystyle\phantom{f^{\xi(2)}_{[ 1, 2, \pm 2]}  =}  +\cos  2 i_{3}) ,  \\
&\displaystyle f^{\xi(2)}_{[ 3,\raisebox{0mm}{-} 2, 0]}  =  \frac{9}{128}  (1-\cos  2  i) (1-\cos  2 i_{3}) , \nonumber \\
&\displaystyle f^{\xi(2)}_{[ 3,\raisebox{0mm}{-} 2, \pm 1]}  =  \frac{3}{64}  (2 \sin   i \pm \sin  2  i) (2 \sin  i_{3} \pm \sin  2 i_{3}) , \nonumber \\
&\displaystyle f^{\xi(2)}_{[ 3,\raisebox{0mm}{-} 2, \pm 2]}  =  \frac{3}{256}  (3 \pm 4 \cos   i+\cos  2  i) (3 \pm 4 \cos  i_{3} \nonumber \\
&\displaystyle\phantom{f^{\xi(2)}_{[ 3,\raisebox{0mm}{-} 2, \pm 2]}  =}    +\cos  2 i_{3}) , \nonumber \\
&\displaystyle f^{\xi(2)}_{[ 3, 0, 0]}  =  \frac{3}{64}  (1-\cos  2  i) (1+3 \cos  2 i_{3}) , \nonumber \\
&\displaystyle f^{\xi(2)}_{[ 3, 0, \pm 1]}  =  -\frac{3}{32}  ( \pm 2 \sin   i+\sin  2  i) \sin  2 i_{3} , \nonumber \\
&\displaystyle f^{\xi(2)}_{[ 3, 0, \pm 2]}  =  \frac{3}{128}  (3 \pm 4 \cos   i+\cos  2  i) (1-\cos  2 i_{3}) , \nonumber \\
&\displaystyle f^{\xi(2)}_{[ 3, 2, 0]}  =  f^{\xi(2)}_{[ 3,\raisebox{0mm}{-} 2, 0]} , \nonumber \\
&\displaystyle f^{\xi(2)}_{[ 3, 2, \pm 1]}  =  -\frac{3}{64}  (2 \sin   i \pm \sin  2  i) (2 \sin  i_{3} \mp \sin  2 i_{3}) , \nonumber \\
&\displaystyle f^{\xi(2)}_{[ 3, 2, \pm 2]}  =  \frac{3}{256}  (3 \pm 4 \cos   i+\cos  2  i) (3 \mp 4 \cos  i_{3}  +\cos  2 i_{3}). \nonumber 
\end{align}

(3) Inclination functions with the form $  f^{\eta (2)}_{[\kappa,\, p,\, q]}(i) $:
\begin{align}
&\displaystyle f^{\eta(2)}_{[ 1,\raisebox{0mm}{-} 2, 0]}  =  -\frac{3}{128}  (11-3 \cos  2  i) (1-\cos  2 i_{3}) , \nonumber \\
&\displaystyle f^{\eta(2)}_{[ 1,\raisebox{0mm}{-} 2, \pm 1]}  =  -\frac{3}{64}  (6 \sin   i \pm \sin  2  i) (2 \sin  i_{3} \pm \sin  2 i_{3}) , \nonumber \\
&\displaystyle f^{\eta(2)}_{[ 1,\raisebox{0mm}{-} 2, \pm 2]}  =  -\frac{3}{256}  (11 \pm 12 \cos   i+\cos  2  i) (3 \pm 4 \cos  i_{3} \nonumber \\
&\displaystyle\phantom{f^{\eta(2)}_{[ 1,\raisebox{0mm}{-} 2, \pm 2]}  =} +\cos  2 i_{3}) , \nonumber \\
&\displaystyle f^{\eta(2)}_{[ 1, 0, 0]}  =  -\frac{1}{64}  (11-3 \cos  2  i) (1+3 \cos  2 i_{3}) , \nonumber \\
&\displaystyle f^{\eta(2)}_{[ 1, 0, \pm 1]}  =  \frac{3}{32}  (\pm 6 \sin   i+\sin  2  i) \sin  2 i_{3} , \nonumber \\
&\displaystyle f^{\eta(2)}_{[ 1, 0, \pm 2]}  =  -\frac{3}{128}  (11 \pm 12 \cos   i+\cos  2  i) (1-\cos  2 i_{3}) , \nonumber \\
&\displaystyle f^{\eta(2)}_{[ 1, 2, 0]}  =  f^{\eta(2)}_{[ 1,\raisebox{0mm}{-} 2, 0]} , \nonumber \\
&\displaystyle f^{\eta(2)}_{[ 1, 2, \pm 1]}  =  \frac{3}{64}  (6 \sin   i \pm \sin  2  i) (2 \sin  i_{3} \mp \sin  2 i_{3}) , \nonumber \\
&\displaystyle f^{\eta(2)}_{[ 1, 2, \pm 2]}  =  -\frac{3}{256}  (11 \pm 12 \cos   i+\cos  2  i) (3 \mp 4 \cos  i_{3} \nonumber \\
&\displaystyle\phantom{f^{\eta(2)}_{[ 1, 2, \pm 2]}  =} +\cos  2 i_{3}) ,  \\
&\displaystyle f^{\eta(2)}_{[ 3,\raisebox{0mm}{-} 2, 0]}  =  \frac{9}{128}  (1-\cos  2  i) (1-\cos  2 i_{3}) , \nonumber \\
&\displaystyle f^{\eta(2)}_{[ 3,\raisebox{0mm}{-} 2, \pm 1]}  =  \frac{3}{64}  (2 \sin   i \pm \sin  2  i) (2 \sin  i_{3} \pm \sin  2 i_{3}) , \nonumber \\
&\displaystyle f^{\eta(2)}_{[ 3,\raisebox{0mm}{-} 2, \pm 2]}  =  \frac{3}{256}  (3 \pm 4 \cos   i+\cos  2  i) (3 \pm 4 \cos  i_{3} \nonumber \\
&\displaystyle\phantom{f^{\eta(2)}_{[ 3,\raisebox{0mm}{-} 2, \pm 2]}  =} +\cos  2 i_{3}) , \nonumber \\
&\displaystyle f^{\eta(2)}_{[ 3, 0, 0]}  =  \frac{3}{64}  (1-\cos  2  i) (1+3 \cos  2 i_{3}) , \nonumber \\
&\displaystyle f^{\eta(2)}_{[ 3, 0, \pm 1]}  =  -\frac{3}{32}  (\pm 2 \sin   i+\sin  2  i) \sin  2 i_{3} , \nonumber \\
&\displaystyle f^{\eta(2)}_{[ 3, 0, \pm 2]}  =  \frac{3}{128}  (3 \pm 4 \cos   i+\cos  2  i) (1-\cos  2 i_{3}) , \nonumber \\
&\displaystyle f^{\eta(2)}_{[ 3, 2, 0]}  =  f^{\eta(2)}_{[ 3,\raisebox{0mm}{-} 2, 0]} , \nonumber \\
&\displaystyle f^{\eta(2)}_{[ 3, 2, \pm 1]}  =  -\frac{3}{64}  (2 \sin   i \pm \sin  2  i) (2 \sin  i_{3} \mp \sin  2 i_{3}) , \nonumber \\
&\displaystyle f^{\eta(2)}_{[ 3, 2, \pm 2]}  =  \frac{3}{256}  (3 \pm 4 \cos   i+\cos  2  i) (3 \mp 4 \cos  i_{3}+\cos  2 i_{3}). \nonumber 
\end{align}

(4) Inclination functions with the form $  f^{i(2)}_{[\kappa,\, p,\, q]}(i) $:
\begin{align}
&  f^{i(2)}_{[ 2,\raisebox{0mm}{-} 2, 0]}  =  \frac{9}{64}  (1-\cos  2 i_{3}) \sin  2  i , \nonumber \\
&  f^{i(2)}_{[ 2,\raisebox{0mm}{-} 2, \pm 1]}  =  \frac{3}{32}  (\cos   i \pm \cos  2  i) (2 \sin  i_{3} \pm \sin  2 i_{3}) , \nonumber \\
&  f^{i(2)}_{[ 2,\raisebox{0mm}{-} 2, \pm 2]}  =  -\frac{3}{128}  (\pm 3+4 \cos  i_{3} \pm \cos  2 i_{3}) (2 \sin   i \pm \sin  2  i) , \nonumber \\
&  f^{i(2)}_{[ 2, 0, 0]}  =  \frac{3}{32}  (1+3 \cos  2 i_{3}) \sin  2  i , \nonumber \\
&  f^{i(2)}_{[ 2, 0, \pm 1]}  =  -\frac{3}{16}  (\pm \cos   i+\cos  2  i) \sin  2 i_{3} ,  \\
&  f^{i(2)}_{[ 2, 0, \pm 2]}  =  -\frac{3}{64}  (1-\cos  2 i_{3}) ( \pm 2 \sin   i+\sin  2  i) , \nonumber \\
&  f^{i(2)}_{[ 2, 2, 0]}  =  f^{i(2)}_{[ 2,\raisebox{0mm}{-} 2, 0]} , \nonumber \\
&  f^{i(2)}_{[ 2, 2, \pm 1]}  =  -\frac{3}{32}  (\cos   i \pm \cos  2  i) (2 \sin  i_{3} \mp \sin  2 i_{3}) , \nonumber \\
&  f^{i(2)}_{[ 2, 2, \pm 2]}  =  -\frac{3}{128}  (\pm 3-4 \cos  i_{3} \pm \cos  2 i_{3}) (2 \sin   i \pm \sin  2  i). \nonumber 
\end{align}

(5) Inclination functions with the form $  f^{\Omega (2)}_{[\kappa,\, p,\, q]}(i) $:
\begin{align}
&  f^{\Omega(2)}_{[ 2,\raisebox{0mm}{-} 2,\raisebox{0mm}{-} 1]}  =  \frac{3}{32}  \sec  \frac{i}{2} \sin  \frac{3 i}{2} (2 \sin  i_{3}-\sin  2 i_{3}) , \nonumber \\
&  f^{\Omega(2)}_{[ 2,\raisebox{0mm}{-} 2, 0]}  =  \frac{9}{32}  \cos   i (1-\cos  2 i_{3}) , \nonumber \\
&  f^{\Omega(2)}_{[ 2,\raisebox{0mm}{-} 2, 1]}  =  \frac{3}{32}  \cos  \frac{3 i}{2} \csc  \frac{i}{2} (2 \sin  i_{3}+\sin  2 i_{3}) , \nonumber \\
&  f^{\Omega(2)}_{[ 2,\raisebox{0mm}{-} 2, \pm 2]}  =  -\frac{3}{64}  (1 \pm \cos   i) (\pm 3+4 \cos  i_{3} \pm \cos  2 i_{3}) , \nonumber \\
&  f^{\Omega(2)}_{[ 2, 0,\raisebox{0mm}{-} 1]}  =  \frac{3}{16}  \sec  \frac{i}{2} \sin  \frac{3 i}{2} \sin  2 i_{3} , \nonumber \\
&  f^{\Omega(2)}_{[ 2, 0, 0]}  =  \frac{3}{16}  \cos   i (1+3 \cos  2 i_{3}) ,  \\
&  f^{\Omega(2)}_{[ 2, 0, 1]}  =  -\frac{3}{16}  \cos  \frac{3 i}{2} \csc  \frac{i}{2} \sin  2 i_{3} , \nonumber \\
&  f^{\Omega(2)}_{[ 2, 0, \pm 2]}  =  -\frac{3}{32}  (\pm 1+\cos   i) (1 - \cos  2 i_{3}) , \nonumber \\
&  f^{\Omega(2)}_{[ 2, 2,\raisebox{0mm}{-} 1]}  =  -\frac{3}{32}  \sec  \frac{i}{2} \sin  \frac{3 i}{2} (2 \sin  i_{3}+\sin  2 i_{3}) , \nonumber \\
&  f^{\Omega(2)}_{[ 2, 2, 0]}  =  f^{\Omega(2)}_{[ 2,\raisebox{0mm}{-} 2, 0]} , \nonumber \\
&  f^{\Omega(2)}_{[ 2, 2, 1]}  =  -\frac{3}{32}  \cos  \frac{3 i}{2} \csc  \frac{i}{2} (2 \sin  i_{3}-\sin  2 i_{3}) , \nonumber \\
&  f^{\Omega(2)}_{[ 2, 2, \pm 2]}  =  -\frac{3}{64}  (\pm 1+\cos   i) (3 \mp 4 \cos  i_{3}+\cos  2 i_{3}).  \nonumber
\end{align}

(6) Inclination functions with the form $  f^{\lambda (2)}_{[\kappa,\, p,\, q]}(i) $:
\begin{align}
&\displaystyle f^{\lambda(2)[a]}_{[ 2,\raisebox{0mm}{-} 2, 0]}  =  -\frac{9}{64}  (1-\cos  2  i) (1-\cos  2 i_{3}) , \nonumber \\
&\displaystyle f^{\lambda(2)[a]}_{[ 2,\raisebox{0mm}{-} 2, \pm 1]}  =  \frac{-3}{32}  (2 \sin   i \pm \sin  2  i) (2 \sin  i_{3} \pm \sin  2 i_{3}) , \nonumber \\
&\displaystyle f^{\lambda(2)[a]}_{[ 2,\raisebox{0mm}{-} 2, \pm 2]}  =  -\frac{3}{128}  (3 \pm 4 \cos   i+\cos  2  i) (3 \pm 4 \cos  i_{3} \nonumber \\
&\displaystyle\phantom{f^{\lambda(2)[a]}_{[ 2,\raisebox{0mm}{-} 2, \pm 2]}  =} +\cos  2 i_{3}) , \nonumber \\
&\displaystyle f^{\lambda(2)[a]}_{[ 2, 0, 0]}  =  -\frac{3}{32}  (1-\cos  2  i) (1+3 \cos  2 i_{3}) , \nonumber \\
&\displaystyle f^{\lambda(2)[a]}_{[ 2, 0, \pm 1]}  =  \frac{3}{16}  (\pm 2 \sin   i+\sin  2  i) \sin  2 i_{3} , \nonumber \\
&\displaystyle f^{\lambda(2)[a]}_{[ 2, 0, \pm 2]}  =  -\frac{3}{64}  (3 \pm 4 \cos   i+\cos  2  i) (1-\cos  2 i_{3}) , \nonumber \\
&\displaystyle f^{\lambda(2)[a]}_{[ 2, 2, 0]}  =  f^{\lambda(2)[a]}_{[ 2,\raisebox{0mm}{-} 2, 0]} , \nonumber \\
&\displaystyle f^{\lambda(2)[a]}_{[ 2, 2, \pm 1]}  =  \frac{3}{32}  (2 \sin   i \pm \sin  2  i) (2 \sin  i_{3} \mp \sin  2 i_{3}) , \nonumber \\
&\displaystyle f^{\lambda(2)[a]}_{[ 2, 2, \pm 2]}  =  -\frac{3}{128}  (3 \pm 4 \cos   i+\cos  2  i) (3 \mp 4 \cos  i_{3} \nonumber \\
&\displaystyle\phantom{f^{\lambda(2)[a]}_{[ 2, 2, \pm 2]}  =} +\cos  2 i_{3}),\nonumber \\
&\displaystyle f^{\lambda(2)[\lambda]}_{[ 2,\raisebox{0mm}{-} 2,\raisebox{0mm}{-} 1]}  =  -\frac{3}{64}  \sec  \frac{i}{2}  (5 \sin  \frac{i}{2}+2 \sin  \frac{3 i}{2}-\sin  \frac{5 i}{2} ) (2 \sin
    i_{3} \nonumber \\
&\displaystyle\phantom{f^{\lambda(2)[\lambda]}_{[ 2,\raisebox{0mm}{-} 2,\raisebox{0mm}{-} 1]}  =} -\sin  2 i_{3}) ,  \\
&\displaystyle f^{\lambda(2)[\lambda]}_{[ 2,\raisebox{0mm}{-} 2, 0]}  =  -\frac{9}{64}  (3-\cos  2  i) (1-\cos  2 i_{3}) , \nonumber \\
&\displaystyle f^{\lambda(2)[\lambda]}_{[ 2,\raisebox{0mm}{-} 2, 1]}  =  -\frac{3}{64}  \csc  \frac{i}{2} (5 \cos  \frac{i}{2}-2 \cos  \frac{3 i}{2}-\cos  \frac{5 i}{2} )  (2 \sin
    i_{3} \nonumber \\
&\displaystyle\phantom{f^{\lambda(2)[\lambda]}_{[ 2,\raisebox{0mm}{-} 2, 1]}  =} +\sin  2 i_{3}) , \nonumber \\
&\displaystyle f^{\lambda(2)[\lambda]}_{[ 2,\raisebox{0mm}{-} 2, \pm 2]}  =  -\frac{3}{128}  (5 \pm 6 \cos   i+\cos  2  i) (3 \pm 4 \cos  i_{3}  \nonumber \\
&\displaystyle\phantom{f^{\lambda(2)[\lambda]}_{[ 2,\raisebox{0mm}{-} 2, \pm 2]}  =}  +\cos  2 i_{3}) , \nonumber \\
&\displaystyle f^{\lambda(2)[\lambda]}_{[ 2, 0,\raisebox{0mm}{-} 1]}  =  -\frac{3}{32}  \sec  \frac{i}{2}  (5 \sin  \frac{i}{2}+2 \sin  \frac{3 i}{2}-\sin  \frac{5 i}{2} ) \sin  2 i_{3} , \nonumber \\
&\displaystyle f^{\lambda(2)[\lambda]}_{[ 2, 0, 0]}  =  -\frac{3}{32}  (3-\cos  2  i) (1+3 \cos  2 i_{3}) , \nonumber \\
&\displaystyle f^{\lambda(2)[\lambda]}_{[ 2, 0, 1]}  =  \frac{3}{32} \csc  \frac{i}{2}   (5 \cos  \frac{i}{2}-2 \cos  \frac{3 i}{2}-\cos  \frac{5 i}{2} ) \sin  2 i_{3} , \nonumber \\
&\displaystyle f^{\lambda(2)[\lambda]}_{[ 2, 0, \pm 2]}  =  -\frac{3}{64}  (5 \pm 6 \cos   i+\cos  2  i) (1-\cos  2 i_{3}) , \nonumber \\
&\displaystyle f^{\lambda(2)[\lambda]}_{[ 2, 2,\raisebox{0mm}{-} 1]}  =  \frac{3}{64}  \sec  \frac{i}{2}  (5 \sin  \frac{i}{2}+2 \sin  \frac{3 i}{2}-\sin  \frac{5 i}{2} ) (2 \sin
    i_{3} \nonumber \\
&\displaystyle\phantom{f^{\lambda(2)[\lambda]}_{[ 2, 2,\raisebox{0mm}{-} 1]}  =} +\sin  2 i_{3}) , \nonumber \\
&\displaystyle f^{\lambda(2)[\lambda]}_{[ 2, 2, 0]}  =  f^{\lambda(2)[\lambda]}_{[ 2,\raisebox{0mm}{-} 2, 0]} , \nonumber \\
&\displaystyle f^{\lambda(2)[\lambda]}_{[ 2, 2, 1]}  =  \frac{3}{64} \csc  \frac{i}{2}   (5 \cos  \frac{i}{2}-2 \cos  \frac{3 i}{2}-\cos  \frac{5 i}{2} ) (2 \sin
    i_{3} \nonumber \\
&\displaystyle\phantom{f^{\lambda(2)[\lambda]}_{[ 2, 2, 1]}  =} -\sin  2 i_{3}) , \nonumber \\
&\displaystyle f^{\lambda(2)[\lambda]}_{[ 2, 2, \pm 2]}  =  \frac{-3}{128}  (5 \pm 6 \cos   i+\cos  2  i) (3 \mp 4 \cos  i_{3}  +\cos  2 i_{3}). \nonumber 
\end{align}


\bibliography{TQorbit}
\end{document}